\documentclass[english,preprint]{revtex4}
\usepackage[latin9]{inputenc}
\usepackage{amstext,bm}
\usepackage{graphicx}
\usepackage{xcolor}
\usepackage{color}
\usepackage{amsmath}
\usepackage{multirow}
\usepackage{babel}
\usepackage[titletoc]{appendix}
\usepackage{hyperref}
\hypersetup{
     colorlinks = true,
     citecolor  = red,  
     linkcolor  = magenta 
}
\usepackage{float}

\newcommand{\kp}{k$\cdot$p }

\begin{document}

\title{Topological Nodal Line Semimetals in CaP$_3$ family of materials}

\author{Qiunan Xu$^{1,2}$}
\author{Rui Yu$^{3}$}
\email{yurui@hit.edu.cn}
\author{Zhong Fang$^{1,2}$}
\author{Xi Dai$^{1,2}$}
\author{Hongming Weng$^{1,2}$}
\email{hmweng@iphy.ac.cn}

\affiliation{$^{1}$Beijing National Laboratory for Condensed Matter Physics,
and Institute of Physics, Chinese Academy of Sciences,
Beijing 100190, China}

\affiliation{$^{2}$Collaborative Innovation Center of Quantum Matter,
Beijing 100190, China}

\affiliation{ $^{3}$Department of Physics, Harbin Institute of Technology,
Harbin 150001, China}

\date{\today}

\begin{abstract}
We propose that CaP$_3$ family of materials,
which include CaP$_{3}$, CaAs$_{3}$, SrP$_{3}$, SrAs$_{3}$ and BaAs$_{3}$
can host a three-dimensional topological nodal line semimetal states.
Based on first-principle calculations and \kp model analysis,
we show that a closed topological nodal line exists near the Fermi energy,
which is protected by the coexistence of time-reversal and spatial inversion
symmetry when the band inversion happens.
A drumhead-like surface states are also obtained on the c-direction surface of these materials.
\end{abstract}

\maketitle

\section{Introduction}
The study of topological semimetals has attracted broad interests
from both the theoretical and the experimental communities in recent years.
Generally speaking, topological semimetals are topologically stable for
their Fermi surfaces enclose nontrivial band crossing points in crystal momentum space.
Such band crossing points behave as the monopoles of
Berry flux~\cite{volovik_book,ZhongFang_monopole_2003},
and bring quantized Berry flux when passing through the surrounding
enclosed Fermi surface~\cite{volovik_book,Weng_ADV:2015hh}.
This quantized number can be taken as the topological invariant to identify the band topology of corresponding metals.
Based on the degeneracy of the band crossing points and its distribution in Brillouin zone, one can classify topological semimetals into
Dirac semimetals, Weyl semimetals and nodal line semimetals.
For Dirac semimetals, the band crossing points are fourfold degenerate,
which can be seen as three dimensional version of Graphene. This novel state has been theoretically proposed and experimentally confirmed in
Na$_3$Bi and Cd$_3$As$_2$ compounds~\cite{Wang:2012ds,Wang:2013is,Yang:2014ia,PhysRevB.91.155139,PhysRevLett.113.246402,Neupane:2014kc,Liu:2014bf,Liu:2014hr}.
For Weyl semimetals, the bands crossing points are double degenerate, with definite chirality and locate at an even number of discrete points in the Brillouin zone, which have been theoretically predicted~\cite{Weng:2015dy,Huang:2015ic,WSM_typeII_BAB} and experimentally verified in TaAs family of materials very recently~\cite{Lv:2015pya,Huang:2015um,Lv:2015kp,Xu07082015,Xu:2015vb,Lv:2015vf}.
For topological nodal line semimetals~\cite{PhysRevB.90.205136,PhysRevB.92.081201}, the band crossing points form closed loops instead of discrete points in the Brillouin zone.
Now many theoretical proposed materials for realizing this topological states have emerged, including
Bernal graphite~\cite{NLS_Heikkila_2011JETP1,NLS_Heikkila_2011JETP2,NLS_Heikkila_2015},
Mackay-Terrones crystal~\cite{NLS_MTC_Weng_PRB},
hyper-honeycomb lattices~\cite{Mullen:2014wq},
Ca$_3$P$_2$~\cite{NLS_Ca3P2_Xie,NLS_Ca3P2_2015_chan},
LaN~\cite{NLS_LaN_FuLiang},
Cu$_3$(Pd,Zn)N~\cite{NLS_Cu3PdN_Yu_PRL,NLS_Cu3PdN_Kane_2015PRL},
the interpenetrated graphene network~\cite{Chen:2015vm},
(Tl,Pb)TaSe$_2$~\cite{NLS_TlTaSe2,NLS_PbTeSe2},
ZrSiS~\cite{NLS_ZrSiS_2015},
perovskite iridates~\cite{NLS_SrIrO_PRB2012,NLS_AIrO3_PRB2015,NLS_SrIrO3_2015,NLS_perovskite_iridates_NC2015},
CaAgX (X=P,As)~\cite{NLS_CaAgX} and
black phosphorus~\cite{NLS_JZZhao_BP}.
The intriguing expected properties characterizing topological nodal line semimetals include the drumhead-like
nearly flat surface states~\cite{Volovik_2011,NLS_MTC_Weng_PRB,NLS_Cu3PdN_Yu_PRL,NLS_Cu3PdN_Kane_2015PRL},
the unique Landau energy level~\cite{NLS_LL_PRB},
long range Coulomb interaction~\cite{NLS_long_range_Coulomb_interaction},
special collective modes~\cite{NLS_Collective_modes_2015} and
opening an important route to achieving
high-temperature superconductivity~\cite{Flatband_HTC_Heikkila_2011PRB,Flatband_HTC_Volovik_2014arXiv,Flatband_HTC_Heikkila_2015arXiv}.

In the present work, based on first-principles calculations and
\kp model Hamiltonian analysis, we predict that CaP$_3$ family of materials
are another candidate for topological nodal line semimetals.
The rest of the paper is organized as follows.
In section II, we present the crystal structure and the
first-principles calculation methodology.
Then we present the calculated bulk and surface electronic structure of
CaP$_3$ family of materials in Sec.~\ref{sec_band}.
In Sec.~\ref{sec_kpmodel}, an effective \kp model is constructed and the nodal line structure and the surface states are studied from the \kp Hamiltonian.
Conclusions are given at the end of this paper.

\section{The crystal structure and computational Method}\label{sec_str}
\begin{figure}[h]
\begin{centering}
\includegraphics[width=1\columnwidth]{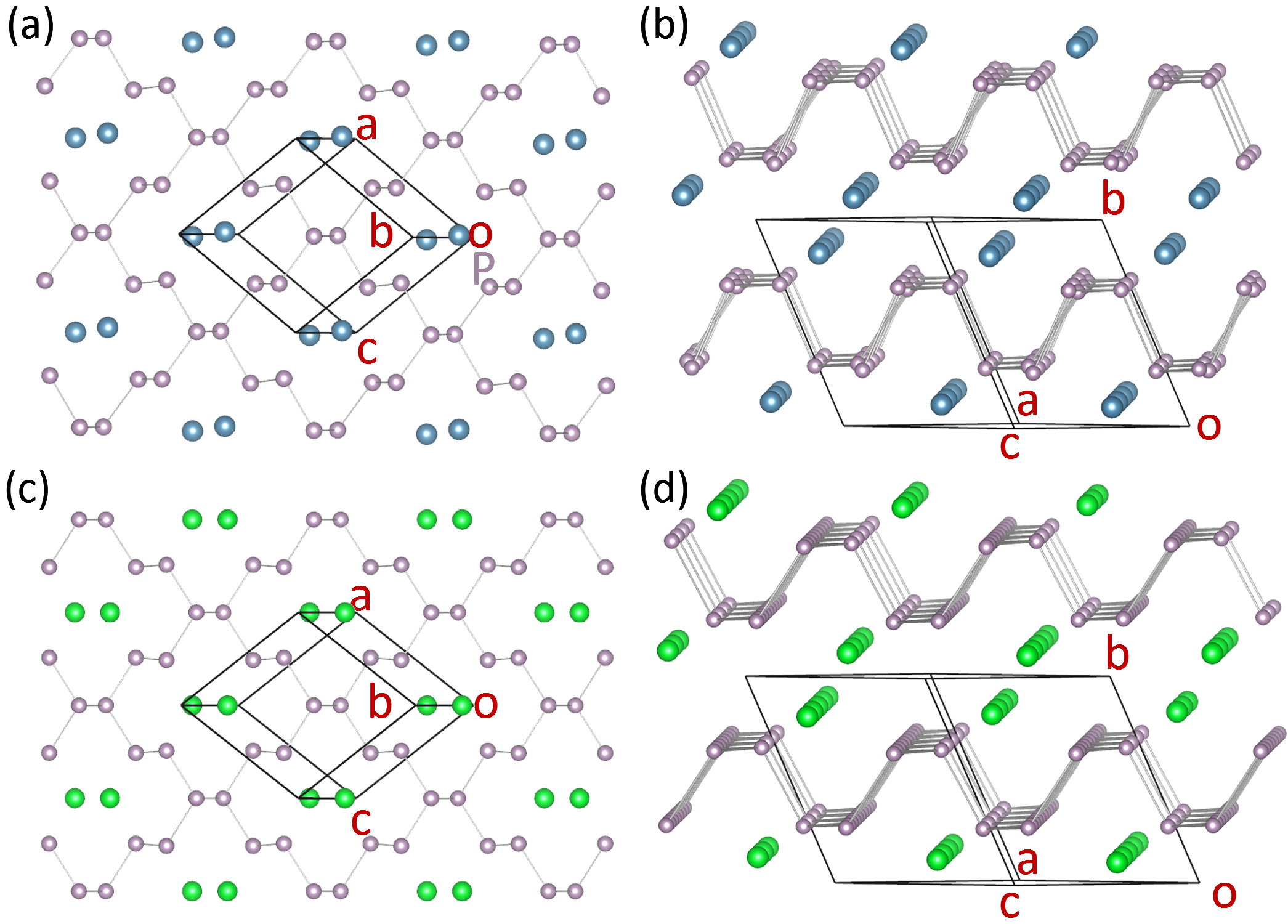}
\par\end{centering}
\protect\caption{\label{fig:CS_BZ}
(Color online)
(a) Top view of single layer of CaP$_3$ and CaAs$_3$.
The puckered P and As layers (gray color balls) are closely related to the orthorhombic black phosphorus and can be derived from the latter by removing 1/4 of the P atoms.
(b) Crystal structure CaP$_3$ and CaAs$_3$. The puckered polyanionic layers stack along the b-axis. The space group for these two compounds is P-1 (No. 2).
(c) and (d) for the single layer and bulk crystal structure of SrP$_{3}$, SrAs$_{3}$ and BaAs$_{3}$, which are characterized by space group C2/m (No. 12).}
\end{figure}

The crystal structure of CaP$_3$ families can be viewed as a list of two dimensional (2D) infinite puckered polyanionic layers
$^2_\infty[\rm P_3]^{2-}$~\cite{CaP3} (see Fig.~\ref{fig:CS_BZ}(a,c))
stacking along the b-axis and forming channels in the a-c direction
with the cations inserting into them as shown in Fig.~\ref{fig:CS_BZ}(b,d).
The space group of CaP$_3$ and CaAs$_3$ is {\it P-1}, while SrP$_3$, SrAs$_3$ and BaAs$_3$ have higher symmetry which is characterized
by space group {\it C2/m}.
The crystallographic data and the atomic coordinates for these materials
are listed in Tab.~\ref{tab:t1} and ~\ref{tab:t2}, and the primitive cell illustrated in Fig.~\ref{fig:CS_BZ} (b,d) are used in the following calculations.
%
%
%
%
%
%
%
%
\begin{table}[H]
\caption{Crystallographic data for CaP$_{3}$, CaAs$_{3}$, SrP$_{3}$, SrAs$_{3}$
and BaAs$_{3}$.}
\begin{centering}
\begin{tabular}{cccccc}
\hline
	Formula & CaP$_{3}$~\cite{CaP3} & CaAs$_{3}$~\cite{CaAs3_BaAs3} & SrP$_{3}$~\cite{SrP3_Chen2003449} & SrAs$_{3}$~\cite{SrAs3} & BaAs$_{3}$~\cite{CaAs3_BaAs3} \tabularnewline
\hline
\multirow{1}{*}{Space group} & {\it P-1} & {\it P-1} & {\it C2/m} & {\it C2/m} & {\it C2/m}\tabularnewline
\textit{a} (nm) & 0.5590 & 0.5866 & 0.7288 & 0.763 & 0.776\tabularnewline
\textit{b} (nm) & 0.5618 & 0.5838 & 0.5690 & 0.588 & 0.6015\tabularnewline
\textit{c} (nm) & 0.5665 & 0.5921 & 0.9199 & 0.961 & 1.0162\tabularnewline
$\alpha$ ($^{\circ}$) & 69.96 & 70.04 & 66.55 & 57.1 & 66.45\tabularnewline
$\beta$ ($^{\circ}$)  & 79.49 & 80.16 & 90 & 90 & 90\tabularnewline
$\gamma$ ($^{\circ}$)  & 74.78 & 75.85 & 90 & 90 & 90\tabularnewline
\hline
\end{tabular}
\par\end{centering}
\label{tab:t1}
\end{table}
%
%
%
%
%
%
%
%
%
%
%
%
%
%
%
%
%
The first-principle calculations are performed by using the Vienna $ab\ initio$
simulation package (VASP) based on generalized gradient approximation (GGA) in
Perdew-Burke-Ernzerhof (PBE)~\cite{Perdew:1996iq} type and the
projector augmented-wave (PAW) pseudo-potential~\cite{Blochl:1994zz}.
The energy cutoff is set to 400 eV for the plane-wave basis and the Brillouin zone
integration was performed on a regular mesh of 8$\times$8$\times$8
k-points. The band structure here is also checked by the nonlocal
Heyd-Scuseria-Ernzerhof (HSE06) hybrid functional calculations.
A tight-binding model based on maximally localized Wannier
functions (MLWF) method \cite{Mostofi:2008ff,Marzari:2012eu} has
been constructed in order to investigate the surface states in the
$c$ direction.
%
%
%
%
%
%
\begin{table}
\caption{Atomic coordinates, equivalent isotropic displacement parameters
for CaP$_{3}$, CaAs$_{3}$, SrP$_{3}$, SrAs$_{3}$ and BaAs$_{3}$.}
\begin{centering}
\begin{tabular}{ccccccc}
\hline
\multirow{2}{*}{Atoms} & \multirow{2}{*}{Site} & Wyckoff & \multirow{2}{*}{Symmetry} & \multirow{2}{*}{x} & \multirow{2}{*}{y} & \multirow{2}{*}{z}\tabularnewline
 &  & symbol &  &  &  & \tabularnewline
\hline
CaP$_{3}$ & Ca & 2i & l  &    0.175	 &  0.141  & 0.146\tabularnewline
 & P1 & 2i & l           &    0.498  &  0.303  &  0.501          \tabularnewline
 & P2 & 2i & l           &    0.630	 &  0.320  & 0.104          \tabularnewline
 & P3 & 2i & l           &    0.104	 &  0.300  & 0.600          \tabularnewline
CaAs$_{3}$ & Ca & 2i & l &    0.1847 &  0.1595 & 0.1243\tabularnewline
 & As1 & 2i & l          &    0.5104 &  0.2829 & 0.5149         \tabularnewline
 & As2 & 2i & l          &    0.6449 &  0.3130 & 0.0928         \tabularnewline
 & As3 & 2i & l          &    0.0895 &  0.2801 & 0.5907         \tabularnewline
SrP$_{3}$ & Sr & 4i & m  &    0    	 &  0.8526 & 0.8382\tabularnewline
 & P1 & 4i & m           &    0	     &  0.6906 & 0.5105         \tabularnewline
 & P2 & 8j & l           &    0.744	 &  0.6795 & 0.6404         \tabularnewline
SrAs$_{3}$ & Sr & 4i & m &    0	     &  0.8345 & 0.8369\tabularnewline
 & As1 & 4i & m          &    0	     &  0.7122 & 0.4960         \tabularnewline
 & As2 & 8j & l          &    0.7676 &  0.6946 & 0.1385         \tabularnewline
BaAs$_{3}$ & Ba & 4i & m &    0	     &  0.8370 & 0.8363\tabularnewline
 & As1 & 4i & m          &    0	     &  0.2969 & 0.4958         \tabularnewline
 & As2 & 8j & l          &    0.7658 &  0.3165 & 0.8614         \tabularnewline
\hline
\end{tabular}
\par\end{centering}
\label{tab:t2}
\end{table}
%
%
%
%
%
%
%
%
\section{Results and Discussion}
\subsection{Electronic structure}\label{sec_band}
\begin{figure}[h]
\begin{centering}
\includegraphics[width=0.751\columnwidth]{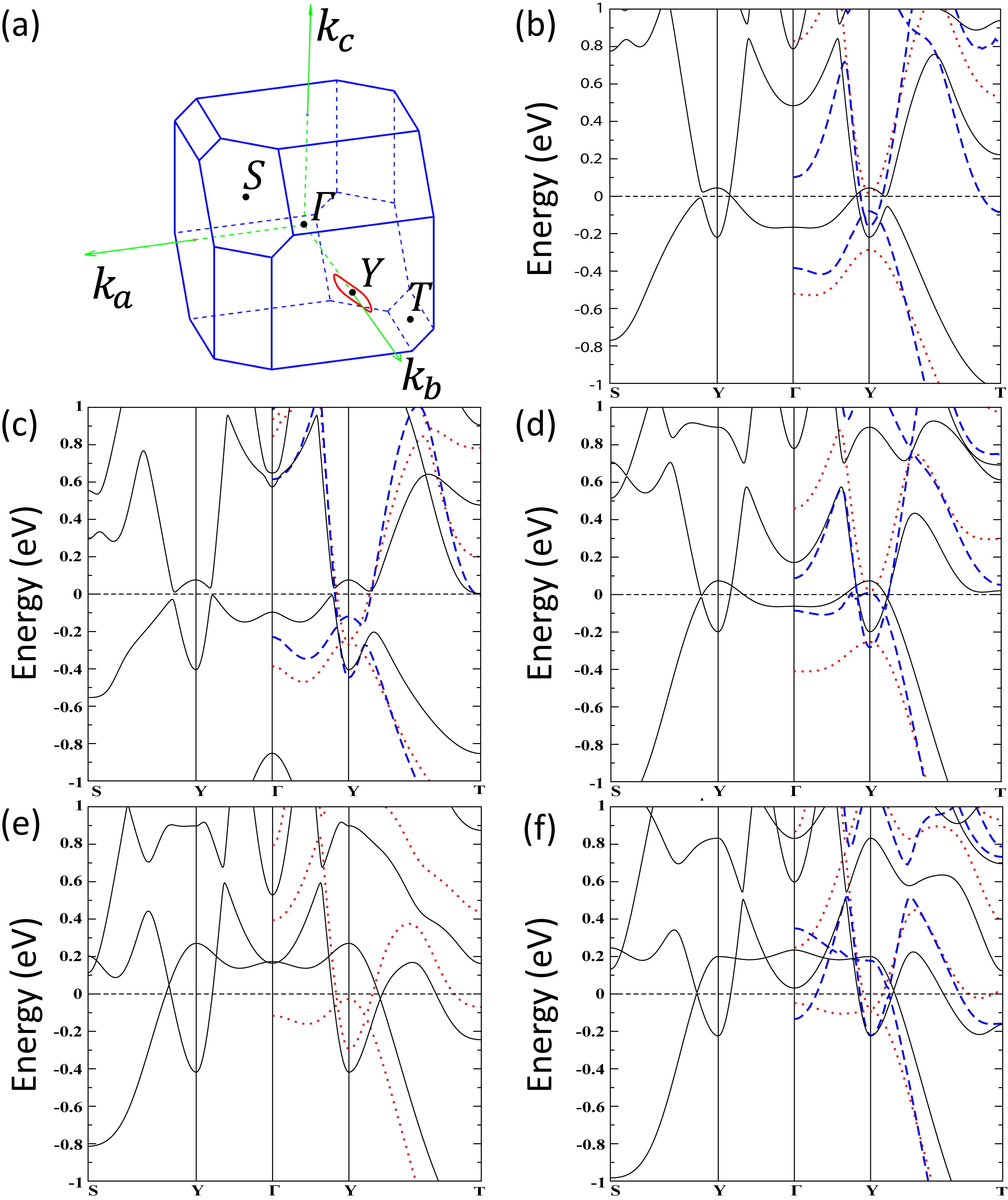}
\par\end{centering}
\protect\caption{\label{fig:bulk_band}
(Color online)
(a) The bulk Brillouin zone for CaP$_3$ family of materials.
The nodal line (red color loop) surrounds $Y$ point and lies on the $\Gamma$-$Y$-$S$ plane for SrP$_3$, SrAs$_3$ and BaAs$_3$ compounds, while it slightly deviates from this plane for CaP$_3$ and CaAs$_3$.
The band structure from GGA calculations are shown with black solid cures for
(b) CaP$_3$, (c) CaAs$_3$, (d)  SrP$_3$, (e) SrAs$_3$ and (f) BaAs$_3$.
The red dotted curves are HSE06 calculation results and the blue dashed curves
are HSE06 calculation results with compressed lattice structure $0.94a\times0.94b\times0.94c$.
}
\end{figure}

The band structure of CaP$_3$ family of materials are presented in
Fig.~\ref{fig:bulk_band} where the spin-orbit coupling does not take into consideration.
The band structure which is got by GGA calculations show that
two bands with opposite parity are inverted around the $Y$ point near the Fermi energy.
The symmetry at the $Y$ point are composed of time-reversal symmetry and space inversion symmetry for {\it P-1} symmetry materials and with an additional mirror symmetry for {\it C2/m} symmetry materials.
As proposed in our early work, in the case of coexistence of time-reversal symmetry and space inversion symmetry, the energy inverted bands with opposite parity should cross along a closed nodal line~\cite{NLS_MTC_Weng_PRB,NLS_Cu3PdN_Yu_PRL,NLS_JZZhao_BP}.
The nodal line structures are found lying on the $\Gamma$-$Y$-$S$ plane for SrP$_3$, SrAs$_3$ and BaAs$_3$ as shown in Fig.~\ref{fig:bulk_band}(a), while for CaP$_3$ and CaAs$_3$ the
nodal line is slightly deviate from this plane.
We have performed the HSE06 calculations to check the band structure near $Y$ point, which is shown by the red doted curves in Fig.~\ref{fig:bulk_band}(b-f).
We find that only SrAs$_3$ takes band inverted structure in the HSE06 calculations, while the band energies of the other four compounds are in normal order.
Therefore the nodal line structure is survived in the former one, but vanished in the latter four materials.
On the other hands, we also find that compressing the lattice volume is benefit for the emerging of the band inversion.
The band structures of compressed lattice with HSE06 calculation are shown with
blue dashed curves in Fig.~\ref{fig:bulk_band}.
Therefore the emergence of the nodal line in the latter four materials can be controlled by compressing the crystal lattice.
%
%
%

%
%
%
\begin{figure}[h]
\begin{centering}
\includegraphics[width=.71\columnwidth]{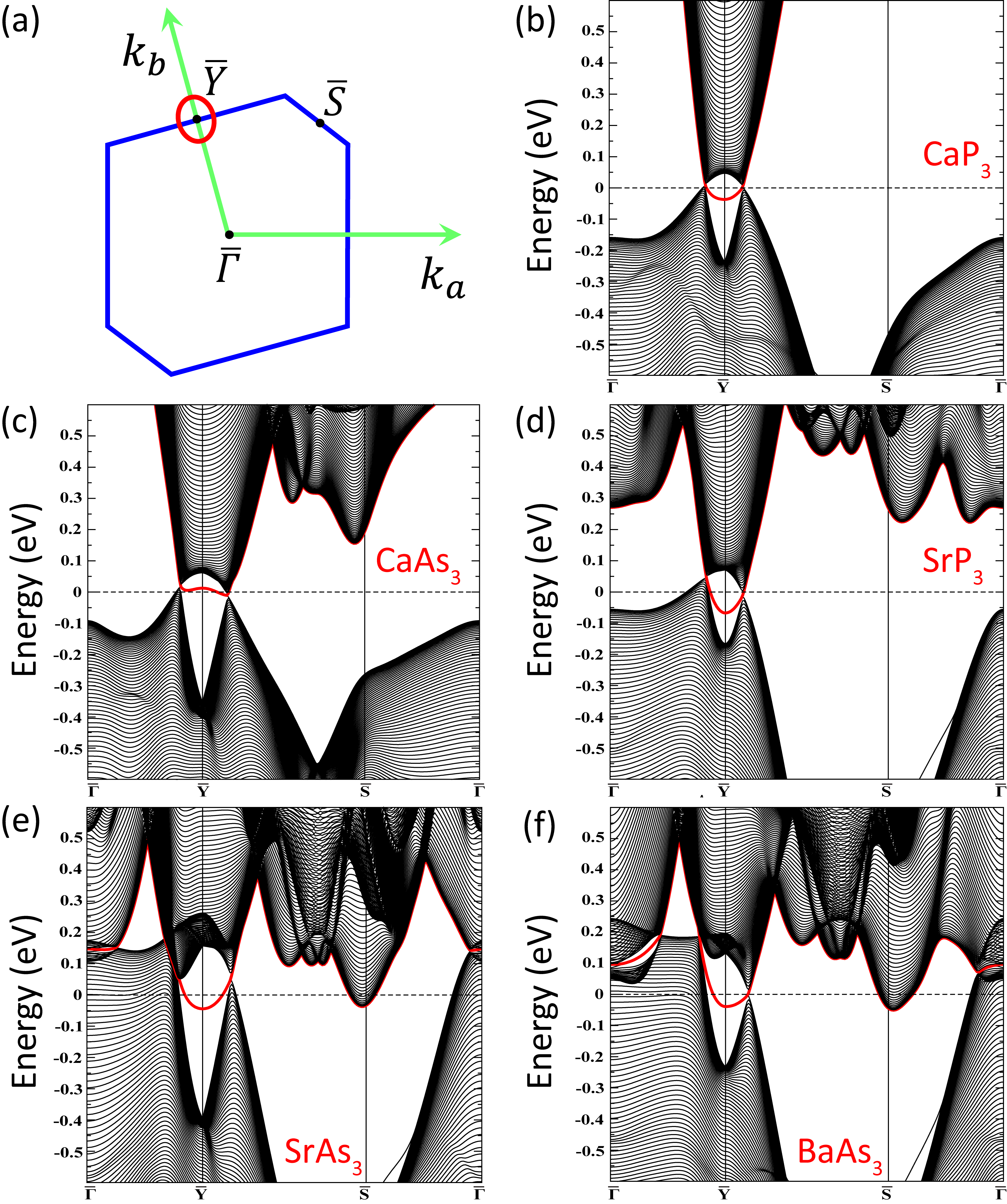}
\par\end{centering}
\protect\caption{\label{fig:surface_states}
(Color online)(a) The projected Brillouin zone along c-direction.
The surface states (red colored curve near Y point) (b) for CaP$_3$, (c) CaAs$_3$, (d) SrP$_3$, (e) SrAs$_3$ and (f) BaAs$_3$ are nestled between two solid Dirac cones, which are the projection of the nodal line circles.
}
\end{figure}

According to the bulk boundary correspondence, the novel surface states are
expected to appear on the surface of materials.
In order to calculate such surface states, we construct a tight-binding Hamiltonian for a thick slab along the c-direction by using the MLWF method~\cite{Mostofi:2008ff,Marzari:2012eu}.
The obtained surface states are nestled between two solid Dirac cones as
shown in Fig.~\ref{fig:surface_states}(b-f), which are the projection of the nodal line circles in the c-direction.
%
%
%
%
%
%
%
%
%
%
%

In the above calculations, the spin-orbit coupling are set to be vanished, and the nodal line structure can be found in the Brillouin zone.
If we take the spin-orbit effect into consideration, gaps will be opened along the nodal line, and these materials become small gap insulators. The gap values along 
$S-Y$ and $Y-\Gamma$ directions are listed in Talbe. III.

\begin{table}[H]
\caption{The gap values near the nodal line after considering the spin-orbit coupling.}
\begin{centering}
\begin{tabular}{cccccc}
\hline
	  & $S-Y$ & $Y-\Gamma$ \tabularnewline
\hline
\multirow{1}{*}
CaP$_{3}$  & 31.69 meV & 3.73 meV \tabularnewline
CaAs$_{3}$ & 54.47 meV & 39.92 meV \tabularnewline
SrP$_{3}$  & 6.11 meV & 1.76 meV \tabularnewline
SrAs$_{3}$ & 47.14 meV & 6.28 meV \tabularnewline
BaAs$_{3}$ & 38.97 meV & 6.22 meV \tabularnewline
\hline
\end{tabular}
\par\end{centering}
\end{table}

\subsection{Model Hamiltonian}\label{sec_kpmodel}
In this section we investigate the nodal line structure from continuous \kp models. First, we construct the \kp model near band inversion point from the symmetry principles.
Then we calculate the energy dispersions of the surface states by using the obtained \kp Hamiltonian.

The most general form of a two-band model can be written as
\begin{equation}
H(\bm{k})=\sum_{i=0}^{3} g_{i}(\bm{k})\sigma_{i},\label{eq:H_kp}
\end{equation}
where $g_i({\bm k})$ are real functions of ${\bm k}$, $\sigma_0$ is identity matrix
and $\sigma_{1,2,3}$ are Pauli matrices for the space expand by
the two investigated bands near Fermi energy.
At the band inversion point $Y$, the symmetry group is reduced to {\it C$_i$}
for {\it P-1} space group materials and {\it C2h} for {\it C2/m} space group materials.

The {\it C$_i$} group contains time reversal symmetry $\hat{T}$ and space inversion symmetry $\hat{P}$.
For the {\it C2h} group, there is an additional mirror symmetry $M:k_a \leftrightarrow k_c;k_b \rightarrow k_b$.
At $Y$ point, the two inverted bands has opposite parity and then the inversion operator
can be choose as $\hat{P}=\sigma_z$.
The inversion symmetry constrains the Hamiltonian satisfying
\begin{equation}
\hat{P}H({\bm k})\hat{P}^{-1}=H({-\bm k}), \label{eq:tht}
\end{equation}
which lead to that
$g_{0,3}({\bm k})$ are even function of $\bm k$ and
$g_{1,2}({\bm k})$ are odd functions of $\bm k$.
On the other hands, the time-reversal symmetry requires that
\begin{equation}
\hat{T}H({\bm k})\hat{T}^{-1}=H({-\bm k}), \label{eq:php}
\end{equation}
where
$\hat{T}=K$ and $K$ is the complex conjugate operator for the spin less case.
The requirement lead to that
$g_{0,1,3}({\bm k})$ are even and
$g_{2}({\bm k})$ is an odd functions of $\bm k$.
Combining the constraints to $g_i({\bm k})$ from time-reversal and space inversion symmetry, we obtain that $g_{1}(\bm{k})=0$, $g_{0,3}({\bm k})$ are even functions of ${\bm k}$ and $g_{2}({\bm k})$ is an odd function of ${\bm k}$.
Keep up to the lowest order of  ${\bm k}$, we get
\begin{eqnarray}
g_{0}(\bm{k}) & = & a_{0}+a_{1}k_{a}^{2}+a_{2}k_{b}^{2}+a_{3}k_{c}^{2},\nonumber \\
g_{2}(\bm{k}) & = & \alpha k_{a}+\beta k_{b}+\gamma k_{c},\nonumber \\
g_{3}(\bm{k}) & = & m_{0}+m_{1}k_{a}^{2}+m_{2}k_{b}^{2}+m_{3}k_{c}^{2}.\label{eq:g_k}
\end{eqnarray}
For simplicity, the basis vectors in ${\bm k}$ space are choosing as
$k_{a}$, $k_{b}$ and ${k_c}$ as shown in Fig.~\ref{fig:bulk_band}(a).
For the {\it P-1} space group materials CaP$_3$ and CaAs$_3$, the parameters in Eq.~(\ref{eq:H_kp}) are independent.
For the {\it C2/m} space group materials, the mirror symmetry can be chosen as
$\hat{M}=\sigma_z$, and the mirror symmetry gives an additional constraint to Hamiltonian Eq.~(\ref{eq:H_kp})
\begin{equation}
\hat{M}H({k_a, k_b, k_c})\hat{M}^{-1}=H({k_c, k_b, k_a}),\label{eq:mpm}
\end{equation}
which reduce the number
of parameters by requiring that $\beta=0$, $a_1=a_3$, $m_1=m_3$ and $\alpha=-\gamma$ in Eq.~(\ref{eq:g_k}).

\begin{figure}[h]
\includegraphics[width=0.7\columnwidth,angle=0]{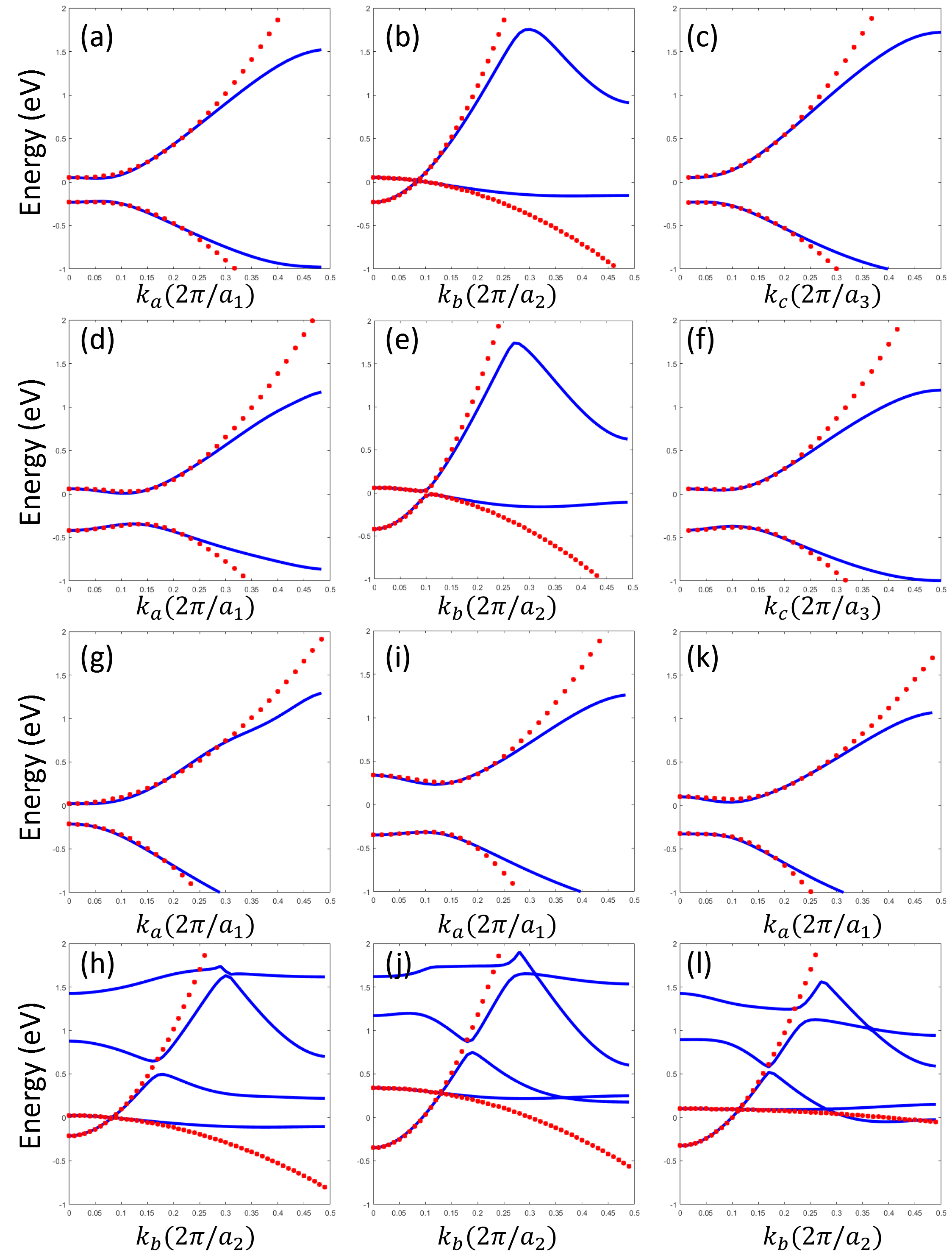}
\caption{\label{fig:kp_bands}
(Color online)
Comparison of band structures of GGA (blue solid curves) and
\kp model (red dashed curves) calculations for (a, b, c) CaP$_3$, (d, e, f) CaAs$_3$, (g, h) SrP$_3$, (i, j) SrAs$_3$ and (k, l) BaAs$_3$.
}
\end{figure}

The \kp parameters obtained by fitting with the first-principle calculations are listed in Tab.~\ref{tab:kp_paras}.
The band structures calculated by the \kp model Hamiltonian are compared with
the first-principle calculations as shown in Fig.~\ref{fig:kp_bands}.
\begin{table}
\caption{The  parameters for \kp Hamiltonian in Eqs.~(\ref{eq:H_kp}) and (\ref{eq:g_k}).
The unit of energy is in eV and the unit of length is in lattice parameters. }\label{tab:kp_paras}
\begin{centering}
\begin{tabular}{c|ccccccc}
\hline
\multirow{4}{*}{CaP$_{3}$}
& $a_{0}$  & $a_{1}$  & $a_{2}$  & $a_{3}$   & $m_{0}$ & $m_{1}$   & $m_{2}$\tabularnewline
& -0.091   &  1.671   & 14.372   & 2.394     & -0.142   & 10.438    & 19.138  \tabularnewline
& $m_{3}$  & $\alpha$ & $\beta$  & $\gamma$  &         &           & \tabularnewline
&  11.910  &  1.773   & 0.001    &  -2.096   & \tabularnewline
\hline
\multirow{4}{*}{CaAs$_{3}$}
& $a_{0}$  & $a_{1}$  & $a_{2}$  & $a_{3}$   & $m_{0}$  & $m_{1}$   & $m_{2}$\tabularnewline
& -0.179   & 1.295    & 17.702   &  1.813    & -0.240   & 9.319     & 23.204 \tabularnewline
& $m_{3}$  & $\alpha$ & $\beta$  & $\gamma$  &          &           & \tabularnewline
& 10.578   & 1.303    & 0.272    & -1.758    & \tabularnewline
\hline
\multirow{2}{*}{SrP$_{3}$}
& $a_{0}$  & $a_{1}$  & $a_{2}$  & $m_{0}$  & $m_{1}$  & $m_{2}$   & $\alpha$\tabularnewline
& -0.095   & -2.324   & 13.619   & -0.1167  & 10.451   & 17.049    & 2.156  \tabularnewline
\hline
\multirow{2}{*}{SrAs$_{3}$ }
& $a_{0}$  & $a_{1}$  & $a_{2}$  & $m_{0}$  & $m_{1}$  & $m_{2}$   & $\alpha$\tabularnewline
& -0.005   & -1.778   & 17.249   & -0.3439  & 12.728   &  20.989   & 1.980  \tabularnewline
\hline
\multirow{2}{*}{BaAs$_{3}$ }
& $a_{0}$  & $a_{1}$  & $a_{2}$  & $m_{0}$  & $m_{1}$  & $m_{2}$   & $\alpha$\tabularnewline
& -0.112   & -3.193   & 15.896   & -0.212   & 11.089   & 16.5431   & 1.918   \tabularnewline
\hline
\end{tabular}
\par\end{centering}
\end{table}

The eigenvalues of Eq.~(\ref{eq:H_kp}) are
$E(\bm{k})=g_{0}(\bm{k})\pm\sqrt{g_{2}^{2}(\bm{k})+g_{3}^{2}(\bm{k})}$ and the
band crossing points appear when $g_{2}(\bm{k})=0$ and $g_{3}(\bm{k})=0$.
In the band inversion case, we obtain that $m_0<0$ and $m_i>0$, $i=1,2,3$.
Then $g_{3}(\bm{k})=m_{0}+m_{1}k_{a}^{2}+m_{2}k_{b}^{2}+m_{3}k_{c}^{2}=0$ is just an equation for an ellipsoidal surface which surrounds $Y$ point in
$\bm{k}$ space.
The second condition $g_{2}(\bm{k})=\alpha k_{a}+\beta k_{b}+\gamma k_{c}=0$
determines a plane passing $Y$ point and with its normal direction along ($\alpha,\beta, \gamma$) direction.
The crossing points between the plane determined by $g_{2}(\bm{k})=0$ and the ellipsoidal surface determined by $g_{3}(\bm{k})=0$ form a closed loop which is just the
band closing nodal line between the two inverted bands.
For the {\it C2/m} space group, where $\beta=0$ and $\alpha=-\gamma$,
the nodal line are calculated lying on the plane that pass through $k_b$ and the angular bisector of $k_a$ and $k_c$.
The higher order terms, such as the fourth order terms in $g_{0,3}(\bm{k})$ and the third order terms in $g_2(\bm{k})$ will deform the ellipsoidal surface and
bending the plane, nevertheless, the crossing nodal line will not disappear but changes to a three dimensional closed loop as shown in Fig.~\ref{fig:CS_BZ}(a).

In the following content, starting from the \kp model in Eqs.~(\ref{eq:H_kp}) and (\ref{eq:g_k}), we present the solutions for the energy
spectra of surface states of CaP$_3$ family of materials.
As shown in Fig.~\ref{fig:CS_BZ}(b), we consider a surface terminated
in $c$ direction. In this case, $k_{c}$ is perpendicular to the surface
and $k_{a,b}$ are parallel to the surface.
Following the method proposed in
Ref. \cite{surface_states_PhysRevB.83.125109}, the
Dirac Hamiltonian in Eq.~(\ref{eq:H_kp}) can be written as
\begin{equation}
H(k_{a},k_{b},k_{c})=g_0({\bm k})+{\bm h (\bm k)}\cdot {\bm \sigma},
\label{eq:H_for_ss1}
\end{equation}
where $\bm \sigma = \left(\sigma_{1}, \sigma_{2}, \sigma_{3}\right) $,
\begin{equation}
{\bm h (\bm k)}=\left[{\bm c}^{0}(k_{a},k_{b})+{\bm c}^{1}(k_{a},k_{b})k_{c}+{\bm c}^{2}(k_{a},k_{b})k_{c}^{2}\right],\label{eq:H_for_ss}
\end{equation}
where
\begin{eqnarray}
\bm{c}^{0} & = & \left(0, \alpha k_{a}+\beta k_{b},m_{0}+m_{1}k_{a}^{2}+m_{2}k_{b}^{2}\right),\nonumber \\
\bm{c}^{1} & = & \left(0, \gamma,0\right),\nonumber \\
\bm{c}^{2} & = & \left(0, 0,m_{3}\right).\label{eq:c_vectors}
\end{eqnarray}
The behavior of ${\bm h (\bm k)}$ completely determines the topological nature of the system
and it is the key to understand the relation between existence of surface states and bulk topological properties.
By tuning $k_c$, the vector ${\bm h (\bm k)}$ forms a parabola in the 2D plane spanned by ${\bm c}^1$ and ${\bm c}^2$.
As proved in Ref. \onlinecite{surface_states_PhysRevB.83.125109}, for the continuum Hamiltonian ${\bm h (\bm k)}$,
the surface states exist if the origin is within the concave side of the parabola,
which lead to the following inequation for  $k_a$ and $k_b$
\begin{equation}
(\alpha k_{a}+\beta k_{b})^2+\frac{\gamma^2}{m_3}(m_{0}+m_{1}k_{a}^{2}+m_{2}k_{b}^{2})<0.
\label{eq:ss01}
\end{equation}
The energy of the surface states (located on the surface of a semi-infinite slab with
$r_c\geq0$) can then be calculated as
\begin{equation}
E_s={\bm c}^0 \cdot \frac{{\bm c}^1\times{\bm c}^2}{|{\bm c}^1\times{\bm c}^2|}.\label{eq:Es}
\end{equation}
As expressed in Eq.~(\ref{eq:c_vectors}),
${\bm c}^0$ is in the plane spanned by vectors ${\bm c}^1$ and ${\bm c}^2$, therefore ${\bm c}^0$ is perpendicular to ${\bm c}^1\times{\bm c}^2$, which leads to $E_s=0$.
This result indicates that a dispersionless state can exist on the surface of a nodal line semimetal within the area determined by Eq.~(\ref{eq:ss01}).
Whereas the topological trivial term $g_0({\bm k})$ in Eq.~(\ref{eq:H_for_ss1}) will introduce a finite dispersion and finally lead to a drumhead-like surface sates as shown in Fig.~\ref{fig:surface_states}.

\section{Conclusion}\label{sec_summary}

In summary, we propose that the 3D topological nodal line semimetal states can be realized in CaP$_3$ family of materials. A closed nodal line is found near the Fermi energy, and this is protected by time-reversal and inversion symmetry with band inverted in the bulk band structure.
The 2D drumhead-like surface states nested inside the closed nodal line are studied in the c-direction by using tight-binding method and \kp model analysis.
Its nearly flat energy dispersion is an ideal playground for many interaction induced nontrivial states, such as fractional topological insulator and high-temperature superconductivity.

\begin{acknowledgments}
This work was supported by the National Natural Science Foundation of China
(No.11274359, No.11422428 and No.41574076), the 973 program of China
(No.2011CBA00108 and No.2013CB921700) and the Strategic Priority Research Program (B) of the
Chinese Academy of Sciences (No.XDB07020100). R.Y. acknowledges funding form
the Fundamental Research Funds for the Central Universities (Grant No. AUGA5710059415) and the National Thousand Young Talents Program.
\end{acknowledgments}

\bibliographystyle{apsrev4-1}
\bibliography{refs}

\begin{thebibliography}{58}%
\makeatletter
\providecommand \@ifxundefined [1]{%
 \@ifx{#1\undefined}
}%
\providecommand \@ifnum [1]{%
 \ifnum #1\expandafter \@firstoftwo
 \else \expandafter \@secondoftwo
 \fi
}%
\providecommand \@ifx [1]{%
 \ifx #1\expandafter \@firstoftwo
 \else \expandafter \@secondoftwo
 \fi
}%
\providecommand \natexlab [1]{#1}%
\providecommand \enquote  [1]{``#1''}%
\providecommand \bibnamefont  [1]{#1}%
\providecommand \bibfnamefont [1]{#1}%
\providecommand \citenamefont [1]{#1}%
\providecommand \href@noop [0]{\@secondoftwo}%
\providecommand \href [0]{\begingroup \@sanitize@url \@href}%
\providecommand \@href[1]{\@@startlink{#1}\@@href}%
\providecommand \@@href[1]{\endgroup#1\@@endlink}%
\providecommand \@sanitize@url [0]{\catcode `\\12\catcode `\$12\catcode
  `\&12\catcode `\#12\catcode `\^12\catcode `\_12\catcode `\%12\relax}%
\providecommand \@@startlink[1]{}%
\providecommand \@@endlink[0]{}%
\providecommand \url  [0]{\begingroup\@sanitize@url \@url }%
\providecommand \@url [1]{\endgroup\@href {#1}{\urlprefix }}%
\providecommand \urlprefix  [0]{URL }%
\providecommand \Eprint [0]{\href }%
\providecommand \doibase [0]{http://dx.doi.org/}%
\providecommand \selectlanguage [0]{\@gobble}%
\providecommand \bibinfo  [0]{\@secondoftwo}%
\providecommand \bibfield  [0]{\@secondoftwo}%
\providecommand \translation [1]{[#1]}%
\providecommand \BibitemOpen [0]{}%
\providecommand \bibitemStop [0]{}%
\providecommand \bibitemNoStop [0]{.\EOS\space}%
\providecommand \EOS [0]{\spacefactor3000\relax}%
\providecommand \BibitemShut  [1]{\csname bibitem#1\endcsname}%
\let\auto@bib@innerbib\@empty
\bibitem [{\citenamefont {Volovik}(2009)}]{volovik_book}%
  \BibitemOpen
  \bibfield  {author} {\bibinfo {author} {\bibfnamefont {G.~E.}\ \bibnamefont
  {Volovik}},\ }\href@noop {} {\emph {\bibinfo {title} {{The Universe in a
  Helium Droplet}}}}\ (\bibinfo  {publisher} {Oxford},\ \bibinfo {year}
  {2009})\BibitemShut {NoStop}%
\bibitem [{\citenamefont {Fang}\ \emph {et~al.}(2003)\citenamefont {Fang},
  \citenamefont {Nagaosa}, \citenamefont {Takahashi}, \citenamefont {Asamitsu},
  \citenamefont {Mathieu}, \citenamefont {Ogasawara}, \citenamefont {Yamada},
  \citenamefont {Kawasaki}, \citenamefont {Tokura},\ and\ \citenamefont
  {Terakura}}]{ZhongFang_monopole_2003}%
  \BibitemOpen
  \bibfield  {author} {\bibinfo {author} {\bibfnamefont {Z.}~\bibnamefont
  {Fang}}, \bibinfo {author} {\bibfnamefont {N.}~\bibnamefont {Nagaosa}},
  \bibinfo {author} {\bibfnamefont {K.~S.}\ \bibnamefont {Takahashi}}, \bibinfo
  {author} {\bibfnamefont {A.}~\bibnamefont {Asamitsu}}, \bibinfo {author}
  {\bibfnamefont {R.}~\bibnamefont {Mathieu}}, \bibinfo {author} {\bibfnamefont
  {T.}~\bibnamefont {Ogasawara}}, \bibinfo {author} {\bibfnamefont
  {H.}~\bibnamefont {Yamada}}, \bibinfo {author} {\bibfnamefont
  {M.}~\bibnamefont {Kawasaki}}, \bibinfo {author} {\bibfnamefont
  {Y.}~\bibnamefont {Tokura}}, \ and\ \bibinfo {author} {\bibfnamefont
  {K.}~\bibnamefont {Terakura}},\ }\href {\doibase 10.1126/science.1089408}
  {\bibfield  {journal} {\bibinfo  {journal} {Science}\ }\textbf {\bibinfo
  {volume} {302}},\ \bibinfo {pages} {92} (\bibinfo {year} {2003})}\BibitemShut
  {NoStop}%
\bibitem [{\citenamefont {Weng}\ \emph
  {et~al.}(2015{\natexlab{a}})\citenamefont {Weng}, \citenamefont {Yu},
  \citenamefont {Hu}, \citenamefont {Dai},\ and\ \citenamefont
  {Fang}}]{Weng_ADV:2015hh}%
  \BibitemOpen
  \bibfield  {author} {\bibinfo {author} {\bibfnamefont {H.}~\bibnamefont
  {Weng}}, \bibinfo {author} {\bibfnamefont {R.}~\bibnamefont {Yu}}, \bibinfo
  {author} {\bibfnamefont {X.}~\bibnamefont {Hu}}, \bibinfo {author}
  {\bibfnamefont {X.}~\bibnamefont {Dai}}, \ and\ \bibinfo {author}
  {\bibfnamefont {Z.}~\bibnamefont {Fang}},\ }\href {\doibase
  10.1080/00018732.2015.1068524} {\bibfield  {journal} {\bibinfo  {journal}
  {Advances in Physics}\ }\textbf {\bibinfo {volume} {64}},\ \bibinfo {pages}
  {227} (\bibinfo {year} {2015}{\natexlab{a}})}\BibitemShut {NoStop}%
\bibitem [{\citenamefont {Wang}\ \emph {et~al.}(2012)\citenamefont {Wang},
  \citenamefont {Sun}, \citenamefont {Chen}, \citenamefont {Franchini},
  \citenamefont {Xu}, \citenamefont {Weng}, \citenamefont {Dai},\ and\
  \citenamefont {Fang}}]{Wang:2012ds}%
  \BibitemOpen
  \bibfield  {author} {\bibinfo {author} {\bibfnamefont {Z.}~\bibnamefont
  {Wang}}, \bibinfo {author} {\bibfnamefont {Y.}~\bibnamefont {Sun}}, \bibinfo
  {author} {\bibfnamefont {X.-Q.}\ \bibnamefont {Chen}}, \bibinfo {author}
  {\bibfnamefont {C.}~\bibnamefont {Franchini}}, \bibinfo {author}
  {\bibfnamefont {G.}~\bibnamefont {Xu}}, \bibinfo {author} {\bibfnamefont
  {H.}~\bibnamefont {Weng}}, \bibinfo {author} {\bibfnamefont {X.}~\bibnamefont
  {Dai}}, \ and\ \bibinfo {author} {\bibfnamefont {Z.}~\bibnamefont {Fang}},\
  }\href {\doibase 10.1103/PhysRevB.85.195320} {\bibfield  {journal} {\bibinfo
  {journal} {Phys. Rev. B}\ }\textbf {\bibinfo {volume} {85}},\ \bibinfo
  {pages} {195320} (\bibinfo {year} {2012})}\BibitemShut {NoStop}%
\bibitem [{\citenamefont {Wang}\ \emph {et~al.}(2013)\citenamefont {Wang},
  \citenamefont {Weng}, \citenamefont {Wu}, \citenamefont {Dai},\ and\
  \citenamefont {Fang}}]{Wang:2013is}%
  \BibitemOpen
  \bibfield  {author} {\bibinfo {author} {\bibfnamefont {Z.}~\bibnamefont
  {Wang}}, \bibinfo {author} {\bibfnamefont {H.}~\bibnamefont {Weng}}, \bibinfo
  {author} {\bibfnamefont {Q.}~\bibnamefont {Wu}}, \bibinfo {author}
  {\bibfnamefont {X.}~\bibnamefont {Dai}}, \ and\ \bibinfo {author}
  {\bibfnamefont {Z.}~\bibnamefont {Fang}},\ }\href {\doibase
  10.1103/PhysRevB.88.125427} {\bibfield  {journal} {\bibinfo  {journal} {Phys.
  Rev. B}\ }\textbf {\bibinfo {volume} {88}},\ \bibinfo {pages} {125427}
  (\bibinfo {year} {2013})}\BibitemShut {NoStop}%
\bibitem [{\citenamefont {Yang}\ and\ \citenamefont
  {Nagaosa}(2014)}]{Yang:2014ia}%
  \BibitemOpen
  \bibfield  {author} {\bibinfo {author} {\bibfnamefont {B.-J.}\ \bibnamefont
  {Yang}}\ and\ \bibinfo {author} {\bibfnamefont {N.}~\bibnamefont {Nagaosa}},\
  }\href {http://dx.doi.org/10.1038/ncomms5898} {\bibfield  {journal} {\bibinfo
   {journal} {Nature Communications}\ }\textbf {\bibinfo {volume} {5}},\
  \bibinfo {pages} {4898} (\bibinfo {year} {2014})}\BibitemShut {NoStop}%
\bibitem [{\citenamefont {Pariari}\ \emph {et~al.}(2015)\citenamefont
  {Pariari}, \citenamefont {Dutta},\ and\ \citenamefont
  {Mandal}}]{PhysRevB.91.155139}%
  \BibitemOpen
  \bibfield  {author} {\bibinfo {author} {\bibfnamefont {A.}~\bibnamefont
  {Pariari}}, \bibinfo {author} {\bibfnamefont {P.}~\bibnamefont {Dutta}}, \
  and\ \bibinfo {author} {\bibfnamefont {P.}~\bibnamefont {Mandal}},\ }\href
  {\doibase 10.1103/PhysRevB.91.155139} {\bibfield  {journal} {\bibinfo
  {journal} {Phys. Rev. B}\ }\textbf {\bibinfo {volume} {91}},\ \bibinfo
  {pages} {155139} (\bibinfo {year} {2015})}\BibitemShut {NoStop}%
\bibitem [{\citenamefont {He}\ \emph {et~al.}(2014)\citenamefont {He},
  \citenamefont {Hong}, \citenamefont {Dong}, \citenamefont {Pan},
  \citenamefont {Zhang}, \citenamefont {Zhang},\ and\ \citenamefont
  {Li}}]{PhysRevLett.113.246402}%
  \BibitemOpen
  \bibfield  {author} {\bibinfo {author} {\bibfnamefont {L.~P.}\ \bibnamefont
  {He}}, \bibinfo {author} {\bibfnamefont {X.~C.}\ \bibnamefont {Hong}},
  \bibinfo {author} {\bibfnamefont {J.~K.}\ \bibnamefont {Dong}}, \bibinfo
  {author} {\bibfnamefont {J.}~\bibnamefont {Pan}}, \bibinfo {author}
  {\bibfnamefont {Z.}~\bibnamefont {Zhang}}, \bibinfo {author} {\bibfnamefont
  {J.}~\bibnamefont {Zhang}}, \ and\ \bibinfo {author} {\bibfnamefont {S.~Y.}\
  \bibnamefont {Li}},\ }\href {\doibase 10.1103/PhysRevLett.113.246402}
  {\bibfield  {journal} {\bibinfo  {journal} {Phys. Rev. Lett.}\ }\textbf
  {\bibinfo {volume} {113}},\ \bibinfo {pages} {246402} (\bibinfo {year}
  {2014})}\BibitemShut {NoStop}%
\bibitem [{\citenamefont {Neupane}\ \emph {et~al.}(2014)\citenamefont
  {Neupane}, \citenamefont {Xu}, \citenamefont {Sankar}, \citenamefont
  {Alidoust}, \citenamefont {Bian}, \citenamefont {Liu}, \citenamefont
  {Belopolski}, \citenamefont {Chang}, \citenamefont {Jeng}, \citenamefont
  {Lin}, \citenamefont {Bansil}, \citenamefont {Chou},\ and\ \citenamefont
  {Hasan}}]{Neupane:2014kc}%
  \BibitemOpen
  \bibfield  {author} {\bibinfo {author} {\bibfnamefont {M.}~\bibnamefont
  {Neupane}}, \bibinfo {author} {\bibfnamefont {S.-Y.}\ \bibnamefont {Xu}},
  \bibinfo {author} {\bibfnamefont {R.}~\bibnamefont {Sankar}}, \bibinfo
  {author} {\bibfnamefont {N.}~\bibnamefont {Alidoust}}, \bibinfo {author}
  {\bibfnamefont {G.}~\bibnamefont {Bian}}, \bibinfo {author} {\bibfnamefont
  {C.}~\bibnamefont {Liu}}, \bibinfo {author} {\bibfnamefont {I.}~\bibnamefont
  {Belopolski}}, \bibinfo {author} {\bibfnamefont {T.-R.}\ \bibnamefont
  {Chang}}, \bibinfo {author} {\bibfnamefont {H.-T.}\ \bibnamefont {Jeng}},
  \bibinfo {author} {\bibfnamefont {H.}~\bibnamefont {Lin}}, \bibinfo {author}
  {\bibfnamefont {A.}~\bibnamefont {Bansil}}, \bibinfo {author} {\bibfnamefont
  {F.}~\bibnamefont {Chou}}, \ and\ \bibinfo {author} {\bibfnamefont {M.~Z.}\
  \bibnamefont {Hasan}},\ }\href {http://dx.doi.org/10.1038/ncomms4786}
  {\bibfield  {journal} {\bibinfo  {journal} {Nature Communications}\ }\textbf
  {\bibinfo {volume} {5}} (\bibinfo {year} {2014})}\BibitemShut {NoStop}%
\bibitem [{\citenamefont {Liu}\ \emph {et~al.}(2014{\natexlab{a}})\citenamefont
  {Liu}, \citenamefont {Zhou}, \citenamefont {Zhang}, \citenamefont {Wang},
  \citenamefont {Weng}, \citenamefont {Prabhakaran}, \citenamefont {Mo},
  \citenamefont {Shen}, \citenamefont {Fang}, \citenamefont {Dai},
  \citenamefont {Hussain},\ and\ \citenamefont {Chen}}]{Liu:2014bf}%
  \BibitemOpen
  \bibfield  {author} {\bibinfo {author} {\bibfnamefont {Z.~K.}\ \bibnamefont
  {Liu}}, \bibinfo {author} {\bibfnamefont {B.}~\bibnamefont {Zhou}}, \bibinfo
  {author} {\bibfnamefont {Y.}~\bibnamefont {Zhang}}, \bibinfo {author}
  {\bibfnamefont {Z.~J.}\ \bibnamefont {Wang}}, \bibinfo {author}
  {\bibfnamefont {H.~M.}\ \bibnamefont {Weng}}, \bibinfo {author}
  {\bibfnamefont {D.}~\bibnamefont {Prabhakaran}}, \bibinfo {author}
  {\bibfnamefont {S.-K.}\ \bibnamefont {Mo}}, \bibinfo {author} {\bibfnamefont
  {Z.~X.}\ \bibnamefont {Shen}}, \bibinfo {author} {\bibfnamefont
  {Z.}~\bibnamefont {Fang}}, \bibinfo {author} {\bibfnamefont {X.}~\bibnamefont
  {Dai}}, \bibinfo {author} {\bibfnamefont {Z.}~\bibnamefont {Hussain}}, \ and\
  \bibinfo {author} {\bibfnamefont {Y.~L.}\ \bibnamefont {Chen}},\ }\href
  {\doibase 10.1126/science.1245085} {\bibfield  {journal} {\bibinfo  {journal}
  {Science}\ }\textbf {\bibinfo {volume} {343}},\ \bibinfo {pages} {864}
  (\bibinfo {year} {2014}{\natexlab{a}})}\BibitemShut {NoStop}%
\bibitem [{\citenamefont {Liu}\ \emph {et~al.}(2014{\natexlab{b}})\citenamefont
  {Liu}, \citenamefont {Jiang}, \citenamefont {Zhou}, \citenamefont {Wang},
  \citenamefont {Zhang}, \citenamefont {Weng}, \citenamefont {Prabhakaran},
  \citenamefont {Mo}, \citenamefont {Peng}, \citenamefont {Dudin},
  \citenamefont {Kim}, \citenamefont {Hoesch}, \citenamefont {Fang},
  \citenamefont {Dai}, \citenamefont {Shen}, \citenamefont {Feng},
  \citenamefont {Hussain},\ and\ \citenamefont {Chen}}]{Liu:2014hr}%
  \BibitemOpen
  \bibfield  {author} {\bibinfo {author} {\bibfnamefont {Z.~K.}\ \bibnamefont
  {Liu}}, \bibinfo {author} {\bibfnamefont {J.}~\bibnamefont {Jiang}}, \bibinfo
  {author} {\bibfnamefont {B.}~\bibnamefont {Zhou}}, \bibinfo {author}
  {\bibfnamefont {Z.~J.}\ \bibnamefont {Wang}}, \bibinfo {author}
  {\bibfnamefont {Y.}~\bibnamefont {Zhang}}, \bibinfo {author} {\bibfnamefont
  {H.~M.}\ \bibnamefont {Weng}}, \bibinfo {author} {\bibfnamefont
  {D.}~\bibnamefont {Prabhakaran}}, \bibinfo {author} {\bibfnamefont {S.~K.}\
  \bibnamefont {Mo}}, \bibinfo {author} {\bibfnamefont {H.}~\bibnamefont
  {Peng}}, \bibinfo {author} {\bibfnamefont {P.}~\bibnamefont {Dudin}},
  \bibinfo {author} {\bibfnamefont {T.}~\bibnamefont {Kim}}, \bibinfo {author}
  {\bibfnamefont {M.}~\bibnamefont {Hoesch}}, \bibinfo {author} {\bibfnamefont
  {Z.}~\bibnamefont {Fang}}, \bibinfo {author} {\bibfnamefont {X.}~\bibnamefont
  {Dai}}, \bibinfo {author} {\bibfnamefont {Z.~X.}\ \bibnamefont {Shen}},
  \bibinfo {author} {\bibfnamefont {D.~L.}\ \bibnamefont {Feng}}, \bibinfo
  {author} {\bibfnamefont {Z.}~\bibnamefont {Hussain}}, \ and\ \bibinfo
  {author} {\bibfnamefont {Y.~L.}\ \bibnamefont {Chen}},\ }\href
  {http://dx.doi.org/10.1038/nmat3990} {\bibfield  {journal} {\bibinfo
  {journal} {Nature Materials}\ }\textbf {\bibinfo {volume} {13}},\ \bibinfo
  {pages} {677} (\bibinfo {year} {2014}{\natexlab{b}})}\BibitemShut {NoStop}%
\bibitem [{\citenamefont {Weng}\ \emph
  {et~al.}(2015{\natexlab{b}})\citenamefont {Weng}, \citenamefont {Fang},
  \citenamefont {Fang}, \citenamefont {Bernevig},\ and\ \citenamefont
  {Dai}}]{Weng:2015dy}%
  \BibitemOpen
  \bibfield  {author} {\bibinfo {author} {\bibfnamefont {H.}~\bibnamefont
  {Weng}}, \bibinfo {author} {\bibfnamefont {C.}~\bibnamefont {Fang}}, \bibinfo
  {author} {\bibfnamefont {Z.}~\bibnamefont {Fang}}, \bibinfo {author}
  {\bibfnamefont {B.~A.}\ \bibnamefont {Bernevig}}, \ and\ \bibinfo {author}
  {\bibfnamefont {X.}~\bibnamefont {Dai}},\ }\href {\doibase
  10.1103/PhysRevX.5.011029} {\bibfield  {journal} {\bibinfo  {journal} {Phys.
  Rev. X}\ }\textbf {\bibinfo {volume} {5}},\ \bibinfo {pages} {011029}
  (\bibinfo {year} {2015}{\natexlab{b}})}\BibitemShut {NoStop}%
\bibitem [{\citenamefont {Huang}\ \emph
  {et~al.}(2015{\natexlab{a}})\citenamefont {Huang}, \citenamefont {Xu},
  \citenamefont {Belopolski}, \citenamefont {Lee}, \citenamefont {Chang},
  \citenamefont {Wang}, \citenamefont {Alidoust}, \citenamefont {Bian},
  \citenamefont {Neupane}, \citenamefont {Zhang}, \citenamefont {Jia},
  \citenamefont {Bansil}, \citenamefont {Lin},\ and\ \citenamefont
  {Hasan}}]{Huang:2015ic}%
  \BibitemOpen
  \bibfield  {author} {\bibinfo {author} {\bibfnamefont {S.-M.}\ \bibnamefont
  {Huang}}, \bibinfo {author} {\bibfnamefont {S.-Y.}\ \bibnamefont {Xu}},
  \bibinfo {author} {\bibfnamefont {I.}~\bibnamefont {Belopolski}}, \bibinfo
  {author} {\bibfnamefont {C.-C.}\ \bibnamefont {Lee}}, \bibinfo {author}
  {\bibfnamefont {G.}~\bibnamefont {Chang}}, \bibinfo {author} {\bibfnamefont
  {B.}~\bibnamefont {Wang}}, \bibinfo {author} {\bibfnamefont {N.}~\bibnamefont
  {Alidoust}}, \bibinfo {author} {\bibfnamefont {G.}~\bibnamefont {Bian}},
  \bibinfo {author} {\bibfnamefont {M.}~\bibnamefont {Neupane}}, \bibinfo
  {author} {\bibfnamefont {C.}~\bibnamefont {Zhang}}, \bibinfo {author}
  {\bibfnamefont {S.}~\bibnamefont {Jia}}, \bibinfo {author} {\bibfnamefont
  {A.}~\bibnamefont {Bansil}}, \bibinfo {author} {\bibfnamefont
  {H.}~\bibnamefont {Lin}}, \ and\ \bibinfo {author} {\bibfnamefont {M.~Z.}\
  \bibnamefont {Hasan}},\ }\href {http://dx.doi.org/10.1038/ncomms8373}
  {\bibfield  {journal} {\bibinfo  {journal} {Nature Communications}\ }\textbf
  {\bibinfo {volume} {6}},\ \bibinfo {pages} {7373} (\bibinfo {year}
  {2015}{\natexlab{a}})}\BibitemShut {NoStop}%
\bibitem [{\citenamefont {Soluyanov}\ \emph {et~al.}(2015)\citenamefont
  {Soluyanov}, \citenamefont {Gresch}, \citenamefont {Wang}, \citenamefont
  {Wu},\ and\ \citenamefont {Troyer}}]{WSM_typeII_BAB}%
  \BibitemOpen
  \bibfield  {author} {\bibinfo {author} {\bibfnamefont {A.~A.}\ \bibnamefont
  {Soluyanov}}, \bibinfo {author} {\bibfnamefont {D.}~\bibnamefont {Gresch}},
  \bibinfo {author} {\bibfnamefont {Z.}~\bibnamefont {Wang}}, \bibinfo {author}
  {\bibfnamefont {Q.~S.}\ \bibnamefont {Wu}}, \ and\ \bibinfo {author}
  {\bibfnamefont {M.}~\bibnamefont {Troyer}},\ }\href@noop {} {\bibfield
  {journal} {\bibinfo  {journal} {Nature}\ }\textbf {\bibinfo {volume} {527}},\
  \bibinfo {pages} {495} (\bibinfo {year} {2015})}\BibitemShut {NoStop}%
\bibitem [{\citenamefont {Lv}\ \emph {et~al.}(2015{\natexlab{a}})\citenamefont
  {Lv}, \citenamefont {Weng}, \citenamefont {Fu}, \citenamefont {Wang},
  \citenamefont {Miao}, \citenamefont {Ma}, \citenamefont {Richard},
  \citenamefont {Huang}, \citenamefont {Zhao}, \citenamefont {Chen},
  \citenamefont {Fang}, \citenamefont {Dai}, \citenamefont {Qian},\ and\
  \citenamefont {Ding}}]{Lv:2015pya}%
  \BibitemOpen
  \bibfield  {author} {\bibinfo {author} {\bibfnamefont {B.~Q.}\ \bibnamefont
  {Lv}}, \bibinfo {author} {\bibfnamefont {H.~M.}\ \bibnamefont {Weng}},
  \bibinfo {author} {\bibfnamefont {B.~B.}\ \bibnamefont {Fu}}, \bibinfo
  {author} {\bibfnamefont {X.~P.}\ \bibnamefont {Wang}}, \bibinfo {author}
  {\bibfnamefont {H.}~\bibnamefont {Miao}}, \bibinfo {author} {\bibfnamefont
  {J.}~\bibnamefont {Ma}}, \bibinfo {author} {\bibfnamefont {P.}~\bibnamefont
  {Richard}}, \bibinfo {author} {\bibfnamefont {X.~C.}\ \bibnamefont {Huang}},
  \bibinfo {author} {\bibfnamefont {L.~X.}\ \bibnamefont {Zhao}}, \bibinfo
  {author} {\bibfnamefont {G.~F.}\ \bibnamefont {Chen}}, \bibinfo {author}
  {\bibfnamefont {Z.}~\bibnamefont {Fang}}, \bibinfo {author} {\bibfnamefont
  {X.}~\bibnamefont {Dai}}, \bibinfo {author} {\bibfnamefont {T.}~\bibnamefont
  {Qian}}, \ and\ \bibinfo {author} {\bibfnamefont {H.}~\bibnamefont {Ding}},\
  }\href {\doibase 10.1103/PhysRevX.5.031013} {\bibfield  {journal} {\bibinfo
  {journal} {Phys. Rev. X}\ }\textbf {\bibinfo {volume} {5}},\ \bibinfo {pages}
  {031013} (\bibinfo {year} {2015}{\natexlab{a}})}\BibitemShut {NoStop}%
\bibitem [{\citenamefont {Huang}\ \emph
  {et~al.}(2015{\natexlab{b}})\citenamefont {Huang}, \citenamefont {Zhao},
  \citenamefont {Long}, \citenamefont {Wang}, \citenamefont {Chen},
  \citenamefont {Yang}, \citenamefont {Liang}, \citenamefont {Xue},
  \citenamefont {Weng}, \citenamefont {Fang}, \citenamefont {Dai},\ and\
  \citenamefont {Chen}}]{Huang:2015um}%
  \BibitemOpen
  \bibfield  {author} {\bibinfo {author} {\bibfnamefont {X.}~\bibnamefont
  {Huang}}, \bibinfo {author} {\bibfnamefont {L.}~\bibnamefont {Zhao}},
  \bibinfo {author} {\bibfnamefont {Y.}~\bibnamefont {Long}}, \bibinfo {author}
  {\bibfnamefont {P.}~\bibnamefont {Wang}}, \bibinfo {author} {\bibfnamefont
  {D.}~\bibnamefont {Chen}}, \bibinfo {author} {\bibfnamefont {Z.}~\bibnamefont
  {Yang}}, \bibinfo {author} {\bibfnamefont {H.}~\bibnamefont {Liang}},
  \bibinfo {author} {\bibfnamefont {M.}~\bibnamefont {Xue}}, \bibinfo {author}
  {\bibfnamefont {H.}~\bibnamefont {Weng}}, \bibinfo {author} {\bibfnamefont
  {Z.}~\bibnamefont {Fang}}, \bibinfo {author} {\bibfnamefont {X.}~\bibnamefont
  {Dai}}, \ and\ \bibinfo {author} {\bibfnamefont {G.}~\bibnamefont {Chen}},\
  }\href {\doibase 10.1103/PhysRevX.5.031023} {\bibfield  {journal} {\bibinfo
  {journal} {Phys. Rev. X}\ }\textbf {\bibinfo {volume} {5}},\ \bibinfo {pages}
  {031023} (\bibinfo {year} {2015}{\natexlab{b}})}\BibitemShut {NoStop}%
\bibitem [{\citenamefont {Lv}\ \emph {et~al.}(2015{\natexlab{b}})\citenamefont
  {Lv}, \citenamefont {Xu}, \citenamefont {Weng}, \citenamefont {Ma},
  \citenamefont {Richard}, \citenamefont {Huang}, \citenamefont {Zhao},
  \citenamefont {Chen}, \citenamefont {Matt}, \citenamefont {Bisti},
  \citenamefont {Strocov}, \citenamefont {Mesot}, \citenamefont {Fang},
  \citenamefont {Dai}, \citenamefont {Qian}, \citenamefont {Shi},\ and\
  \citenamefont {Ding}}]{Lv:2015kp}%
  \BibitemOpen
  \bibfield  {author} {\bibinfo {author} {\bibfnamefont {B.~Q.}\ \bibnamefont
  {Lv}}, \bibinfo {author} {\bibfnamefont {N.}~\bibnamefont {Xu}}, \bibinfo
  {author} {\bibfnamefont {H.~M.}\ \bibnamefont {Weng}}, \bibinfo {author}
  {\bibfnamefont {J.~Z.}\ \bibnamefont {Ma}}, \bibinfo {author} {\bibfnamefont
  {P.}~\bibnamefont {Richard}}, \bibinfo {author} {\bibfnamefont {X.~C.}\
  \bibnamefont {Huang}}, \bibinfo {author} {\bibfnamefont {L.~X.}\ \bibnamefont
  {Zhao}}, \bibinfo {author} {\bibfnamefont {G.~F.}\ \bibnamefont {Chen}},
  \bibinfo {author} {\bibfnamefont {C.~E.}\ \bibnamefont {Matt}}, \bibinfo
  {author} {\bibfnamefont {F.}~\bibnamefont {Bisti}}, \bibinfo {author}
  {\bibfnamefont {V.~N.}\ \bibnamefont {Strocov}}, \bibinfo {author}
  {\bibfnamefont {J.}~\bibnamefont {Mesot}}, \bibinfo {author} {\bibfnamefont
  {Z.}~\bibnamefont {Fang}}, \bibinfo {author} {\bibfnamefont {X.}~\bibnamefont
  {Dai}}, \bibinfo {author} {\bibfnamefont {T.}~\bibnamefont {Qian}}, \bibinfo
  {author} {\bibfnamefont {M.}~\bibnamefont {Shi}}, \ and\ \bibinfo {author}
  {\bibfnamefont {H.}~\bibnamefont {Ding}},\ }\href
  {http://dx.doi.org/10.1038/nphys3426} {\bibfield  {journal} {\bibinfo
  {journal} {Nature Physics}\ }\textbf {\bibinfo {volume} {11}},\ \bibinfo
  {pages} {724} (\bibinfo {year} {2015}{\natexlab{b}})}\BibitemShut {NoStop}%
\bibitem [{\citenamefont {Xu}\ \emph {et~al.}(2015)\citenamefont {Xu},
  \citenamefont {Belopolski}, \citenamefont {Alidoust}, \citenamefont
  {Neupane}, \citenamefont {Bian}, \citenamefont {Zhang}, \citenamefont
  {Sankar}, \citenamefont {Chang}, \citenamefont {Yuan}, \citenamefont {Lee},
  \citenamefont {Huang}, \citenamefont {Zheng}, \citenamefont {Ma},
  \citenamefont {Sanchez}, \citenamefont {Wang}, \citenamefont {Bansil},
  \citenamefont {Chou}, \citenamefont {Shibayev}, \citenamefont {Lin},
  \citenamefont {Jia},\ and\ \citenamefont {Hasan}}]{Xu07082015}%
  \BibitemOpen
  \bibfield  {author} {\bibinfo {author} {\bibfnamefont {S.-Y.}\ \bibnamefont
  {Xu}}, \bibinfo {author} {\bibfnamefont {I.}~\bibnamefont {Belopolski}},
  \bibinfo {author} {\bibfnamefont {N.}~\bibnamefont {Alidoust}}, \bibinfo
  {author} {\bibfnamefont {M.}~\bibnamefont {Neupane}}, \bibinfo {author}
  {\bibfnamefont {G.}~\bibnamefont {Bian}}, \bibinfo {author} {\bibfnamefont
  {C.}~\bibnamefont {Zhang}}, \bibinfo {author} {\bibfnamefont
  {R.}~\bibnamefont {Sankar}}, \bibinfo {author} {\bibfnamefont
  {G.}~\bibnamefont {Chang}}, \bibinfo {author} {\bibfnamefont
  {Z.}~\bibnamefont {Yuan}}, \bibinfo {author} {\bibfnamefont {C.-C.}\
  \bibnamefont {Lee}}, \bibinfo {author} {\bibfnamefont {S.-M.}\ \bibnamefont
  {Huang}}, \bibinfo {author} {\bibfnamefont {H.}~\bibnamefont {Zheng}},
  \bibinfo {author} {\bibfnamefont {J.}~\bibnamefont {Ma}}, \bibinfo {author}
  {\bibfnamefont {D.~S.}\ \bibnamefont {Sanchez}}, \bibinfo {author}
  {\bibfnamefont {B.}~\bibnamefont {Wang}}, \bibinfo {author} {\bibfnamefont
  {A.}~\bibnamefont {Bansil}}, \bibinfo {author} {\bibfnamefont
  {F.}~\bibnamefont {Chou}}, \bibinfo {author} {\bibfnamefont {P.~P.}\
  \bibnamefont {Shibayev}}, \bibinfo {author} {\bibfnamefont {H.}~\bibnamefont
  {Lin}}, \bibinfo {author} {\bibfnamefont {S.}~\bibnamefont {Jia}}, \ and\
  \bibinfo {author} {\bibfnamefont {M.~Z.}\ \bibnamefont {Hasan}},\ }\href
  {\doibase 10.1126/science.aaa9297} {\bibfield  {journal} {\bibinfo  {journal}
  {Science}\ }\textbf {\bibinfo {volume} {349}},\ \bibinfo {pages} {613}
  (\bibinfo {year} {2015})}\BibitemShut {NoStop}%
\bibitem [{\citenamefont {{Xu}}\ \emph {et~al.}(2016)\citenamefont {{Xu}},
  \citenamefont {{Weng}}, \citenamefont {{Lv}}, \citenamefont {{Matt}},
  \citenamefont {{Park}}, \citenamefont {{Bisti}}, \citenamefont {{Strocov}},
  \citenamefont {{Gawryluk}}, \citenamefont {{Pomjakushina}}, \citenamefont
  {{Conder}}, \citenamefont {{Plumb}}, \citenamefont {{Radovic}}, \citenamefont
  {{Aut{\`e}s}}, \citenamefont {{Yazyev}}, \citenamefont {{Fang}},
  \citenamefont {{Dai}}, \citenamefont {{Qian}}, \citenamefont {{Mesot}},
  \citenamefont {{Ding}},\ and\ \citenamefont {{Shi}}}]{Xu:2015vb}%
  \BibitemOpen
  \bibfield  {author} {\bibinfo {author} {\bibfnamefont {N.}~\bibnamefont
  {{Xu}}}, \bibinfo {author} {\bibfnamefont {H.~M.}\ \bibnamefont {{Weng}}},
  \bibinfo {author} {\bibfnamefont {B.~Q.}\ \bibnamefont {{Lv}}}, \bibinfo
  {author} {\bibfnamefont {C.~E.}\ \bibnamefont {{Matt}}}, \bibinfo {author}
  {\bibfnamefont {J.}~\bibnamefont {{Park}}}, \bibinfo {author} {\bibfnamefont
  {F.}~\bibnamefont {{Bisti}}}, \bibinfo {author} {\bibfnamefont {V.~N.}\
  \bibnamefont {{Strocov}}}, \bibinfo {author} {\bibfnamefont {D.}~\bibnamefont
  {{Gawryluk}}}, \bibinfo {author} {\bibfnamefont {E.}~\bibnamefont
  {{Pomjakushina}}}, \bibinfo {author} {\bibfnamefont {K.}~\bibnamefont
  {{Conder}}}, \bibinfo {author} {\bibfnamefont {N.~C.}\ \bibnamefont
  {{Plumb}}}, \bibinfo {author} {\bibfnamefont {M.}~\bibnamefont {{Radovic}}},
  \bibinfo {author} {\bibfnamefont {G.}~\bibnamefont {{Aut{\`e}s}}}, \bibinfo
  {author} {\bibfnamefont {O.~V.}\ \bibnamefont {{Yazyev}}}, \bibinfo {author}
  {\bibfnamefont {Z.}~\bibnamefont {{Fang}}}, \bibinfo {author} {\bibfnamefont
  {X.}~\bibnamefont {{Dai}}}, \bibinfo {author} {\bibfnamefont
  {T.}~\bibnamefont {{Qian}}}, \bibinfo {author} {\bibfnamefont
  {J.}~\bibnamefont {{Mesot}}}, \bibinfo {author} {\bibfnamefont
  {H.}~\bibnamefont {{Ding}}}, \ and\ \bibinfo {author} {\bibfnamefont
  {M.}~\bibnamefont {{Shi}}},\ }\href {\doibase 10.1038/ncomms11006} {\bibfield
   {journal} {\bibinfo  {journal} {Nature Communications}\ }\textbf {\bibinfo
  {volume} {7}},\ \bibinfo {eid} {11006} (\bibinfo {year} {2016})}\BibitemShut
  {NoStop}%
\bibitem [{\citenamefont {Lv}\ \emph {et~al.}(2015{\natexlab{c}})\citenamefont
  {Lv}, \citenamefont {Muff}, \citenamefont {Qian}, \citenamefont {Song},
  \citenamefont {Nie}, \citenamefont {Xu}, \citenamefont {Richard},
  \citenamefont {Matt}, \citenamefont {Plumb}, \citenamefont {Zhao},
  \citenamefont {Chen}, \citenamefont {Fang}, \citenamefont {Dai},
  \citenamefont {Dil}, \citenamefont {Mesot}, \citenamefont {Shi},
  \citenamefont {Weng},\ and\ \citenamefont {Ding}}]{Lv:2015vf}%
  \BibitemOpen
  \bibfield  {author} {\bibinfo {author} {\bibfnamefont {B.~Q.}\ \bibnamefont
  {Lv}}, \bibinfo {author} {\bibfnamefont {S.}~\bibnamefont {Muff}}, \bibinfo
  {author} {\bibfnamefont {T.}~\bibnamefont {Qian}}, \bibinfo {author}
  {\bibfnamefont {Z.~D.}\ \bibnamefont {Song}}, \bibinfo {author}
  {\bibfnamefont {S.~M.}\ \bibnamefont {Nie}}, \bibinfo {author} {\bibfnamefont
  {N.}~\bibnamefont {Xu}}, \bibinfo {author} {\bibfnamefont {P.}~\bibnamefont
  {Richard}}, \bibinfo {author} {\bibfnamefont {C.~E.}\ \bibnamefont {Matt}},
  \bibinfo {author} {\bibfnamefont {N.~C.}\ \bibnamefont {Plumb}}, \bibinfo
  {author} {\bibfnamefont {L.~X.}\ \bibnamefont {Zhao}}, \bibinfo {author}
  {\bibfnamefont {G.~F.}\ \bibnamefont {Chen}}, \bibinfo {author}
  {\bibfnamefont {Z.}~\bibnamefont {Fang}}, \bibinfo {author} {\bibfnamefont
  {X.}~\bibnamefont {Dai}}, \bibinfo {author} {\bibfnamefont {J.~H.}\
  \bibnamefont {Dil}}, \bibinfo {author} {\bibfnamefont {J.}~\bibnamefont
  {Mesot}}, \bibinfo {author} {\bibfnamefont {M.}~\bibnamefont {Shi}}, \bibinfo
  {author} {\bibfnamefont {H.~M.}\ \bibnamefont {Weng}}, \ and\ \bibinfo
  {author} {\bibfnamefont {H.}~\bibnamefont {Ding}},\ }\href {\doibase
  10.1103/PhysRevLett.115.217601} {\bibfield  {journal} {\bibinfo  {journal}
  {Phys. Rev. Lett.}\ }\textbf {\bibinfo {volume} {115}},\ \bibinfo {pages}
  {217601} (\bibinfo {year} {2015}{\natexlab{c}})}\BibitemShut {NoStop}%
\bibitem [{\citenamefont {Chiu}\ and\ \citenamefont
  {Schnyder}(2014)}]{PhysRevB.90.205136}%
  \BibitemOpen
  \bibfield  {author} {\bibinfo {author} {\bibfnamefont {C.-K.}\ \bibnamefont
  {Chiu}}\ and\ \bibinfo {author} {\bibfnamefont {A.~P.}\ \bibnamefont
  {Schnyder}},\ }\href {\doibase 10.1103/PhysRevB.90.205136} {\bibfield
  {journal} {\bibinfo  {journal} {Phys. Rev. B}\ }\textbf {\bibinfo {volume}
  {90}},\ \bibinfo {pages} {205136} (\bibinfo {year} {2014})}\BibitemShut
  {NoStop}%
\bibitem [{\citenamefont {Fang}\ \emph {et~al.}(2015)\citenamefont {Fang},
  \citenamefont {Chen}, \citenamefont {Kee},\ and\ \citenamefont
  {Fu}}]{PhysRevB.92.081201}%
  \BibitemOpen
  \bibfield  {author} {\bibinfo {author} {\bibfnamefont {C.}~\bibnamefont
  {Fang}}, \bibinfo {author} {\bibfnamefont {Y.}~\bibnamefont {Chen}}, \bibinfo
  {author} {\bibfnamefont {H.-Y.}\ \bibnamefont {Kee}}, \ and\ \bibinfo
  {author} {\bibfnamefont {L.}~\bibnamefont {Fu}},\ }\href {\doibase
  10.1103/PhysRevB.92.081201} {\bibfield  {journal} {\bibinfo  {journal} {Phys.
  Rev. B}\ }\textbf {\bibinfo {volume} {92}},\ \bibinfo {pages} {081201}
  (\bibinfo {year} {2015})}\BibitemShut {NoStop}%
\bibitem [{\citenamefont {Heikkil{\"a}}\ \emph {et~al.}(2011)\citenamefont
  {Heikkil{\"a}}, \citenamefont {Kopnin},\ and\ \citenamefont
  {Volovik}}]{NLS_Heikkila_2011JETP1}%
  \BibitemOpen
  \bibfield  {author} {\bibinfo {author} {\bibfnamefont {T.~T.}\ \bibnamefont
  {Heikkil{\"a}}}, \bibinfo {author} {\bibfnamefont {N.~B.}\ \bibnamefont
  {Kopnin}}, \ and\ \bibinfo {author} {\bibfnamefont {G.~E.}\ \bibnamefont
  {Volovik}},\ }\href
  {http://link.springer.com/article/10.1134%2FS0021364011150045} {\bibfield
  {journal} {\bibinfo  {journal} {JETP Letters}\ }\textbf {\bibinfo {volume}
  {94}},\ \bibinfo {pages} {233} (\bibinfo {year} {2011})}\BibitemShut
  {NoStop}%
\bibitem [{\citenamefont {Heikkil{\"a}}\ and\ \citenamefont
  {Volovik}(2011)}]{NLS_Heikkila_2011JETP2}%
  \BibitemOpen
  \bibfield  {author} {\bibinfo {author} {\bibfnamefont {T.~T.}\ \bibnamefont
  {Heikkil{\"a}}}\ and\ \bibinfo {author} {\bibfnamefont {G.~E.}\ \bibnamefont
  {Volovik}},\ }\href
  {http://link.springer.com/article/10.1134%2FS002136401102007X} {\bibfield
  {journal} {\bibinfo  {journal} {JETP Letters}\ }\textbf {\bibinfo {volume}
  {93}},\ \bibinfo {pages} {59} (\bibinfo {year} {2011})}\BibitemShut {NoStop}%
\bibitem [{\citenamefont {{Heikkila}}\ and\ \citenamefont
  {{Volovik}}(2015{\natexlab{a}})}]{NLS_Heikkila_2015}%
  \BibitemOpen
  \bibfield  {author} {\bibinfo {author} {\bibfnamefont {T.~T.}\ \bibnamefont
  {{Heikkila}}}\ and\ \bibinfo {author} {\bibfnamefont {G.~E.}\ \bibnamefont
  {{Volovik}}},\ }\href@noop {} {\bibfield  {journal} {\bibinfo  {journal}
  {ArXiv e-prints}\ } (\bibinfo {year} {2015}{\natexlab{a}})},\ \Eprint
  {http://arxiv.org/abs/1504.05824} {arXiv:1504.05824 [cond-mat.mtrl-sci]}
  \BibitemShut {NoStop}%
\bibitem [{\citenamefont {Weng}\ \emph
  {et~al.}(2015{\natexlab{c}})\citenamefont {Weng}, \citenamefont {Liang},
  \citenamefont {Xu}, \citenamefont {Yu}, \citenamefont {Fang}, \citenamefont
  {Dai},\ and\ \citenamefont {Kawazoe}}]{NLS_MTC_Weng_PRB}%
  \BibitemOpen
  \bibfield  {author} {\bibinfo {author} {\bibfnamefont {H.}~\bibnamefont
  {Weng}}, \bibinfo {author} {\bibfnamefont {Y.}~\bibnamefont {Liang}},
  \bibinfo {author} {\bibfnamefont {Q.}~\bibnamefont {Xu}}, \bibinfo {author}
  {\bibfnamefont {R.}~\bibnamefont {Yu}}, \bibinfo {author} {\bibfnamefont
  {Z.}~\bibnamefont {Fang}}, \bibinfo {author} {\bibfnamefont {X.}~\bibnamefont
  {Dai}}, \ and\ \bibinfo {author} {\bibfnamefont {Y.}~\bibnamefont
  {Kawazoe}},\ }\href {\doibase 10.1103/PhysRevB.92.045108} {\bibfield
  {journal} {\bibinfo  {journal} {Phys. Rev. B}\ }\textbf {\bibinfo {volume}
  {92}},\ \bibinfo {pages} {045108} (\bibinfo {year}
  {2015}{\natexlab{c}})}\BibitemShut {NoStop}%
\bibitem [{\citenamefont {Mullen}\ \emph {et~al.}(2015)\citenamefont {Mullen},
  \citenamefont {Uchoa},\ and\ \citenamefont {Glatzhofer}}]{Mullen:2014wq}%
  \BibitemOpen
  \bibfield  {author} {\bibinfo {author} {\bibfnamefont {K.}~\bibnamefont
  {Mullen}}, \bibinfo {author} {\bibfnamefont {B.}~\bibnamefont {Uchoa}}, \
  and\ \bibinfo {author} {\bibfnamefont {D.~T.}\ \bibnamefont {Glatzhofer}},\
  }\href {\doibase 10.1103/PhysRevLett.115.026403} {\bibfield  {journal}
  {\bibinfo  {journal} {Phys. Rev. Lett.}\ }\textbf {\bibinfo {volume} {115}},\
  \bibinfo {pages} {026403} (\bibinfo {year} {2015})}\BibitemShut {NoStop}%
\bibitem [{\citenamefont {Xie}\ \emph {et~al.}(2015)\citenamefont {Xie},
  \citenamefont {Schoop}, \citenamefont {Seibel}, \citenamefont {Gibson},
  \citenamefont {Xie},\ and\ \citenamefont {Cava}}]{NLS_Ca3P2_Xie}%
  \BibitemOpen
  \bibfield  {author} {\bibinfo {author} {\bibfnamefont {L.~S.}\ \bibnamefont
  {Xie}}, \bibinfo {author} {\bibfnamefont {L.~M.}\ \bibnamefont {Schoop}},
  \bibinfo {author} {\bibfnamefont {E.~M.}\ \bibnamefont {Seibel}}, \bibinfo
  {author} {\bibfnamefont {Q.~D.}\ \bibnamefont {Gibson}}, \bibinfo {author}
  {\bibfnamefont {W.}~\bibnamefont {Xie}}, \ and\ \bibinfo {author}
  {\bibfnamefont {R.~J.}\ \bibnamefont {Cava}},\ }\href {\doibase
  http://dx.doi.org/10.1063/1.4926545} {\bibfield  {journal} {\bibinfo
  {journal} {APL Materials}\ }\textbf {\bibinfo {volume} {3}},\ \bibinfo {eid}
  {083602} (\bibinfo {year} {2015})}\BibitemShut {NoStop}%
\bibitem [{\citenamefont {Chan}\ \emph {et~al.}(2016)\citenamefont {Chan},
  \citenamefont {Chiu}, \citenamefont {Chou},\ and\ \citenamefont
  {Schnyder}}]{NLS_Ca3P2_2015_chan}%
  \BibitemOpen
  \bibfield  {author} {\bibinfo {author} {\bibfnamefont {Y.-H.}\ \bibnamefont
  {Chan}}, \bibinfo {author} {\bibfnamefont {C.-K.}\ \bibnamefont {Chiu}},
  \bibinfo {author} {\bibfnamefont {M.~Y.}\ \bibnamefont {Chou}}, \ and\
  \bibinfo {author} {\bibfnamefont {A.~P.}\ \bibnamefont {Schnyder}},\ }\href
  {\doibase 10.1103/PhysRevB.93.205132} {\bibfield  {journal} {\bibinfo
  {journal} {Phys. Rev. B}\ }\textbf {\bibinfo {volume} {93}},\ \bibinfo
  {pages} {205132} (\bibinfo {year} {2016})}\BibitemShut {NoStop}%
\bibitem [{\citenamefont {{Zeng}}\ \emph {et~al.}(2015)\citenamefont {{Zeng}},
  \citenamefont {{Fang}}, \citenamefont {{Chang}}, \citenamefont {{Chen}},
  \citenamefont {{Hsieh}}, \citenamefont {{Bansil}}, \citenamefont {{Lin}},\
  and\ \citenamefont {{Fu}}}]{NLS_LaN_FuLiang}%
  \BibitemOpen
  \bibfield  {author} {\bibinfo {author} {\bibfnamefont {M.}~\bibnamefont
  {{Zeng}}}, \bibinfo {author} {\bibfnamefont {C.}~\bibnamefont {{Fang}}},
  \bibinfo {author} {\bibfnamefont {G.}~\bibnamefont {{Chang}}}, \bibinfo
  {author} {\bibfnamefont {Y.-A.}\ \bibnamefont {{Chen}}}, \bibinfo {author}
  {\bibfnamefont {T.}~\bibnamefont {{Hsieh}}}, \bibinfo {author} {\bibfnamefont
  {A.}~\bibnamefont {{Bansil}}}, \bibinfo {author} {\bibfnamefont
  {H.}~\bibnamefont {{Lin}}}, \ and\ \bibinfo {author} {\bibfnamefont
  {L.}~\bibnamefont {{Fu}}},\ }\href@noop {} {\bibfield  {journal} {\bibinfo
  {journal} {ArXiv e-prints}\ } (\bibinfo {year} {2015})},\ \Eprint
  {http://arxiv.org/abs/1504.03492} {arXiv:1504.03492 [cond-mat.mes-hall]}
  \BibitemShut {NoStop}%
\bibitem [{\citenamefont {Yu}\ \emph {et~al.}(2015)\citenamefont {Yu},
  \citenamefont {Weng}, \citenamefont {Fang}, \citenamefont {Dai},\ and\
  \citenamefont {Hu}}]{NLS_Cu3PdN_Yu_PRL}%
  \BibitemOpen
  \bibfield  {author} {\bibinfo {author} {\bibfnamefont {R.}~\bibnamefont
  {Yu}}, \bibinfo {author} {\bibfnamefont {H.}~\bibnamefont {Weng}}, \bibinfo
  {author} {\bibfnamefont {Z.}~\bibnamefont {Fang}}, \bibinfo {author}
  {\bibfnamefont {X.}~\bibnamefont {Dai}}, \ and\ \bibinfo {author}
  {\bibfnamefont {X.}~\bibnamefont {Hu}},\ }\href {\doibase
  10.1103/PhysRevLett.115.036807} {\bibfield  {journal} {\bibinfo  {journal}
  {Phys. Rev. Lett.}\ }\textbf {\bibinfo {volume} {115}},\ \bibinfo {pages}
  {036807} (\bibinfo {year} {2015})}\BibitemShut {NoStop}%
\bibitem [{\citenamefont {Kim}\ \emph {et~al.}(2015{\natexlab{a}})\citenamefont
  {Kim}, \citenamefont {Wieder}, \citenamefont {Kane},\ and\ \citenamefont
  {Rappe}}]{NLS_Cu3PdN_Kane_2015PRL}%
  \BibitemOpen
  \bibfield  {author} {\bibinfo {author} {\bibfnamefont {Y.}~\bibnamefont
  {Kim}}, \bibinfo {author} {\bibfnamefont {B.~J.}\ \bibnamefont {Wieder}},
  \bibinfo {author} {\bibfnamefont {C.~L.}\ \bibnamefont {Kane}}, \ and\
  \bibinfo {author} {\bibfnamefont {A.~M.}\ \bibnamefont {Rappe}},\ }\href
  {\doibase 10.1103/PhysRevLett.115.036806} {\bibfield  {journal} {\bibinfo
  {journal} {Phys. Rev. Lett.}\ }\textbf {\bibinfo {volume} {115}},\ \bibinfo
  {pages} {036806} (\bibinfo {year} {2015}{\natexlab{a}})}\BibitemShut
  {NoStop}%
\bibitem [{\citenamefont {Chen}\ \emph
  {et~al.}(2015{\natexlab{a}})\citenamefont {Chen}, \citenamefont {Xie},
  \citenamefont {Yang}, \citenamefont {Pan}, \citenamefont {Zhang},
  \citenamefont {Cohen},\ and\ \citenamefont {zhang}}]{Chen:2015vm}%
  \BibitemOpen
  \bibfield  {author} {\bibinfo {author} {\bibfnamefont {Y.}~\bibnamefont
  {Chen}}, \bibinfo {author} {\bibfnamefont {Y.}~\bibnamefont {Xie}}, \bibinfo
  {author} {\bibfnamefont {S.~A.}\ \bibnamefont {Yang}}, \bibinfo {author}
  {\bibfnamefont {H.}~\bibnamefont {Pan}}, \bibinfo {author} {\bibfnamefont
  {F.}~\bibnamefont {Zhang}}, \bibinfo {author} {\bibfnamefont {M.~L.}\
  \bibnamefont {Cohen}}, \ and\ \bibinfo {author} {\bibfnamefont
  {s.}~\bibnamefont {zhang}},\ }\href@noop {} {\bibfield  {journal} {\bibinfo
  {journal} {Nano Letters}\ }\textbf {\bibinfo {volume} {15}},\ \bibinfo
  {pages} {6974} (\bibinfo {year} {2015}{\natexlab{a}})}\BibitemShut {NoStop}%
\bibitem [{\citenamefont {Bian}\ \emph {et~al.}(2016)\citenamefont {Bian},
  \citenamefont {Chang}, \citenamefont {Zheng}, \citenamefont {Velury},
  \citenamefont {Xu}, \citenamefont {Neupert}, \citenamefont {Chiu},
  \citenamefont {Huang}, \citenamefont {Sanchez}, \citenamefont {Belopolski},
  \citenamefont {Alidoust}, \citenamefont {Chen}, \citenamefont {Chang},
  \citenamefont {Bansil}, \citenamefont {Jeng}, \citenamefont {Lin},\ and\
  \citenamefont {Hasan}}]{NLS_TlTaSe2}%
  \BibitemOpen
  \bibfield  {author} {\bibinfo {author} {\bibfnamefont {G.}~\bibnamefont
  {Bian}}, \bibinfo {author} {\bibfnamefont {T.-R.}\ \bibnamefont {Chang}},
  \bibinfo {author} {\bibfnamefont {H.}~\bibnamefont {Zheng}}, \bibinfo
  {author} {\bibfnamefont {S.}~\bibnamefont {Velury}}, \bibinfo {author}
  {\bibfnamefont {S.-Y.}\ \bibnamefont {Xu}}, \bibinfo {author} {\bibfnamefont
  {T.}~\bibnamefont {Neupert}}, \bibinfo {author} {\bibfnamefont {C.-K.}\
  \bibnamefont {Chiu}}, \bibinfo {author} {\bibfnamefont {S.-M.}\ \bibnamefont
  {Huang}}, \bibinfo {author} {\bibfnamefont {D.~S.}\ \bibnamefont {Sanchez}},
  \bibinfo {author} {\bibfnamefont {I.}~\bibnamefont {Belopolski}}, \bibinfo
  {author} {\bibfnamefont {N.}~\bibnamefont {Alidoust}}, \bibinfo {author}
  {\bibfnamefont {P.-J.}\ \bibnamefont {Chen}}, \bibinfo {author}
  {\bibfnamefont {G.}~\bibnamefont {Chang}}, \bibinfo {author} {\bibfnamefont
  {A.}~\bibnamefont {Bansil}}, \bibinfo {author} {\bibfnamefont {H.-T.}\
  \bibnamefont {Jeng}}, \bibinfo {author} {\bibfnamefont {H.}~\bibnamefont
  {Lin}}, \ and\ \bibinfo {author} {\bibfnamefont {M.~Z.}\ \bibnamefont
  {Hasan}},\ }\href {\doibase 10.1103/PhysRevB.93.121113} {\bibfield  {journal}
  {\bibinfo  {journal} {Phys. Rev. B}\ }\textbf {\bibinfo {volume} {93}},\
  \bibinfo {pages} {121113} (\bibinfo {year} {2016})}\BibitemShut {NoStop}%
\bibitem [{\citenamefont {{Bian}}\ \emph {et~al.}(2015)\citenamefont {{Bian}},
  \citenamefont {{Chang}}, \citenamefont {{Sankar}}, \citenamefont {{Xu}},
  \citenamefont {{Zheng}}, \citenamefont {{Neupert}}, \citenamefont {{Chiu}},
  \citenamefont {{Huang}}, \citenamefont {{Chang}}, \citenamefont
  {{Belopolski}}, \citenamefont {{Sanchez}}, \citenamefont {{Neupane}},
  \citenamefont {{Alidoust}}, \citenamefont {{Liu}}, \citenamefont {{Wang}},
  \citenamefont {{Lee}}, \citenamefont {{Jeng}}, \citenamefont {{Bansil}},
  \citenamefont {{Chou}}, \citenamefont {{Lin}},\ and\ \citenamefont {{Zahid
  Hasan}}}]{NLS_PbTeSe2}%
  \BibitemOpen
  \bibfield  {author} {\bibinfo {author} {\bibfnamefont {G.}~\bibnamefont
  {{Bian}}}, \bibinfo {author} {\bibfnamefont {T.-R.}\ \bibnamefont {{Chang}}},
  \bibinfo {author} {\bibfnamefont {R.}~\bibnamefont {{Sankar}}}, \bibinfo
  {author} {\bibfnamefont {S.-Y.}\ \bibnamefont {{Xu}}}, \bibinfo {author}
  {\bibfnamefont {H.}~\bibnamefont {{Zheng}}}, \bibinfo {author} {\bibfnamefont
  {T.}~\bibnamefont {{Neupert}}}, \bibinfo {author} {\bibfnamefont {C.-K.}\
  \bibnamefont {{Chiu}}}, \bibinfo {author} {\bibfnamefont {S.-M.}\
  \bibnamefont {{Huang}}}, \bibinfo {author} {\bibfnamefont {G.}~\bibnamefont
  {{Chang}}}, \bibinfo {author} {\bibfnamefont {I.}~\bibnamefont
  {{Belopolski}}}, \bibinfo {author} {\bibfnamefont {D.~S.}\ \bibnamefont
  {{Sanchez}}}, \bibinfo {author} {\bibfnamefont {M.}~\bibnamefont
  {{Neupane}}}, \bibinfo {author} {\bibfnamefont {N.}~\bibnamefont
  {{Alidoust}}}, \bibinfo {author} {\bibfnamefont {C.}~\bibnamefont {{Liu}}},
  \bibinfo {author} {\bibfnamefont {B.}~\bibnamefont {{Wang}}}, \bibinfo
  {author} {\bibfnamefont {C.-C.}\ \bibnamefont {{Lee}}}, \bibinfo {author}
  {\bibfnamefont {H.-T.}\ \bibnamefont {{Jeng}}}, \bibinfo {author}
  {\bibfnamefont {A.}~\bibnamefont {{Bansil}}}, \bibinfo {author}
  {\bibfnamefont {F.}~\bibnamefont {{Chou}}}, \bibinfo {author} {\bibfnamefont
  {H.}~\bibnamefont {{Lin}}}, \ and\ \bibinfo {author} {\bibfnamefont
  {M.}~\bibnamefont {{Zahid Hasan}}},\ }\href@noop {} {\bibfield  {journal}
  {\bibinfo  {journal} {ArXiv e-prints}\ } (\bibinfo {year} {2015})},\ \Eprint
  {http://arxiv.org/abs/1505.03069} {arXiv:1505.03069 [cond-mat.mes-hall]}
  \BibitemShut {NoStop}%
\bibitem [{\citenamefont {{Schoop}}\ \emph {et~al.}(2016)\citenamefont
  {{Schoop}}, \citenamefont {{Ali}}, \citenamefont {{Stra{\ss}er}},
  \citenamefont {{Topp}}, \citenamefont {{Varykhalov}}, \citenamefont
  {{Marchenko}}, \citenamefont {{Duppel}}, \citenamefont {{Parkin}},
  \citenamefont {{Lotsch}},\ and\ \citenamefont {{Ast}}}]{NLS_ZrSiS_2015}%
  \BibitemOpen
  \bibfield  {author} {\bibinfo {author} {\bibfnamefont {L.~M.}\ \bibnamefont
  {{Schoop}}}, \bibinfo {author} {\bibfnamefont {M.~N.}\ \bibnamefont {{Ali}}},
  \bibinfo {author} {\bibfnamefont {C.}~\bibnamefont {{Stra{\ss}er}}}, \bibinfo
  {author} {\bibfnamefont {A.}~\bibnamefont {{Topp}}}, \bibinfo {author}
  {\bibfnamefont {A.}~\bibnamefont {{Varykhalov}}}, \bibinfo {author}
  {\bibfnamefont {D.}~\bibnamefont {{Marchenko}}}, \bibinfo {author}
  {\bibfnamefont {V.}~\bibnamefont {{Duppel}}}, \bibinfo {author}
  {\bibfnamefont {S.~S.~P.}\ \bibnamefont {{Parkin}}}, \bibinfo {author}
  {\bibfnamefont {B.~V.}\ \bibnamefont {{Lotsch}}}, \ and\ \bibinfo {author}
  {\bibfnamefont {C.~R.}\ \bibnamefont {{Ast}}},\ }\href
  {http://dx.doi.org/10.1038/ncomms11696} {\bibfield  {journal} {\bibinfo
  {journal} {Nature Communications}\ }\textbf {\bibinfo {volume} {7}},\
  \bibinfo {pages} {11696} (\bibinfo {year} {2016})}\BibitemShut {NoStop}%
\bibitem [{\citenamefont {Carter}\ \emph {et~al.}(2012)\citenamefont {Carter},
  \citenamefont {Shankar}, \citenamefont {Zeb},\ and\ \citenamefont
  {Kee}}]{NLS_SrIrO_PRB2012}%
  \BibitemOpen
  \bibfield  {author} {\bibinfo {author} {\bibfnamefont {J.-M.}\ \bibnamefont
  {Carter}}, \bibinfo {author} {\bibfnamefont {V.~V.}\ \bibnamefont {Shankar}},
  \bibinfo {author} {\bibfnamefont {M.~A.}\ \bibnamefont {Zeb}}, \ and\
  \bibinfo {author} {\bibfnamefont {H.-Y.}\ \bibnamefont {Kee}},\ }\href
  {\doibase 10.1103/PhysRevB.85.115105} {\bibfield  {journal} {\bibinfo
  {journal} {Phys. Rev. B}\ }\textbf {\bibinfo {volume} {85}},\ \bibinfo
  {pages} {115105} (\bibinfo {year} {2012})}\BibitemShut {NoStop}%
\bibitem [{\citenamefont {Kim}\ \emph {et~al.}(2015{\natexlab{b}})\citenamefont
  {Kim}, \citenamefont {Chen},\ and\ \citenamefont {Kee}}]{NLS_AIrO3_PRB2015}%
  \BibitemOpen
  \bibfield  {author} {\bibinfo {author} {\bibfnamefont {H.-S.}\ \bibnamefont
  {Kim}}, \bibinfo {author} {\bibfnamefont {Y.}~\bibnamefont {Chen}}, \ and\
  \bibinfo {author} {\bibfnamefont {H.-Y.}\ \bibnamefont {Kee}},\ }\href
  {\doibase 10.1103/PhysRevB.91.235103} {\bibfield  {journal} {\bibinfo
  {journal} {Phys. Rev. B}\ }\textbf {\bibinfo {volume} {91}},\ \bibinfo
  {pages} {235103} (\bibinfo {year} {2015}{\natexlab{b}})}\BibitemShut
  {NoStop}%
\bibitem [{\citenamefont {Liu}\ \emph {et~al.}(2016)\citenamefont {Liu},
  \citenamefont {Kriegner}, \citenamefont {Horak}, \citenamefont {Puggioni},
  \citenamefont {Rayan~Serrao}, \citenamefont {Chen}, \citenamefont {Yi},
  \citenamefont {Frontera}, \citenamefont {Holy}, \citenamefont {Vishwanath},
  \citenamefont {Rondinelli}, \citenamefont {Marti},\ and\ \citenamefont
  {Ramesh}}]{NLS_SrIrO3_2015}%
  \BibitemOpen
  \bibfield  {author} {\bibinfo {author} {\bibfnamefont {J.}~\bibnamefont
  {Liu}}, \bibinfo {author} {\bibfnamefont {D.}~\bibnamefont {Kriegner}},
  \bibinfo {author} {\bibfnamefont {L.}~\bibnamefont {Horak}}, \bibinfo
  {author} {\bibfnamefont {D.}~\bibnamefont {Puggioni}}, \bibinfo {author}
  {\bibfnamefont {C.}~\bibnamefont {Rayan~Serrao}}, \bibinfo {author}
  {\bibfnamefont {R.}~\bibnamefont {Chen}}, \bibinfo {author} {\bibfnamefont
  {D.}~\bibnamefont {Yi}}, \bibinfo {author} {\bibfnamefont {C.}~\bibnamefont
  {Frontera}}, \bibinfo {author} {\bibfnamefont {V.}~\bibnamefont {Holy}},
  \bibinfo {author} {\bibfnamefont {A.}~\bibnamefont {Vishwanath}}, \bibinfo
  {author} {\bibfnamefont {J.~M.}\ \bibnamefont {Rondinelli}}, \bibinfo
  {author} {\bibfnamefont {X.}~\bibnamefont {Marti}}, \ and\ \bibinfo {author}
  {\bibfnamefont {R.}~\bibnamefont {Ramesh}},\ }\href {\doibase
  10.1103/PhysRevB.93.085118} {\bibfield  {journal} {\bibinfo  {journal} {Phys.
  Rev. B}\ }\textbf {\bibinfo {volume} {93}},\ \bibinfo {pages} {085118}
  (\bibinfo {year} {2016})}\BibitemShut {NoStop}%
\bibitem [{\citenamefont {Chen}\ \emph
  {et~al.}(2015{\natexlab{b}})\citenamefont {Chen}, \citenamefont {Lu},\ and\
  \citenamefont {Kee}}]{NLS_perovskite_iridates_NC2015}%
  \BibitemOpen
  \bibfield  {author} {\bibinfo {author} {\bibfnamefont {Y.}~\bibnamefont
  {Chen}}, \bibinfo {author} {\bibfnamefont {Y.~M.}\ \bibnamefont {Lu}}, \ and\
  \bibinfo {author} {\bibfnamefont {H.~Y.}\ \bibnamefont {Kee}},\ }\href
  {http://www.nature.com/doifinder/10.1038/ncomms7593} {\bibfield  {journal}
  {\bibinfo  {journal} {Nature Communications}\ }\textbf {\bibinfo {volume}
  {6}},\ \bibinfo {pages} {7593} (\bibinfo {year}
  {2015}{\natexlab{b}})}\BibitemShut {NoStop}%
\bibitem [{\citenamefont {Yamakage}\ \emph {et~al.}(2016)\citenamefont
  {Yamakage}, \citenamefont {Yamakawa}, \citenamefont {Tanaka},\ and\
  \citenamefont {Okamoto}}]{NLS_CaAgX}%
  \BibitemOpen
  \bibfield  {author} {\bibinfo {author} {\bibfnamefont {A.}~\bibnamefont
  {Yamakage}}, \bibinfo {author} {\bibfnamefont {Y.}~\bibnamefont {Yamakawa}},
  \bibinfo {author} {\bibfnamefont {Y.}~\bibnamefont {Tanaka}}, \ and\ \bibinfo
  {author} {\bibfnamefont {Y.}~\bibnamefont {Okamoto}},\ }\href {\doibase
  10.7566/JPSJ.85.013708} {\bibfield  {journal} {\bibinfo  {journal} {JPSJ}\
  }\textbf {\bibinfo {volume} {85}},\ \bibinfo {pages} {013708} (\bibinfo
  {year} {2016})}\BibitemShut {NoStop}%
\bibitem [{\citenamefont {{Zhao}}\ \emph {et~al.}(2015)\citenamefont {{Zhao}},
  \citenamefont {{Yu}}, \citenamefont {{Weng}},\ and\ \citenamefont
  {{Fang}}}]{NLS_JZZhao_BP}%
  \BibitemOpen
  \bibfield  {author} {\bibinfo {author} {\bibfnamefont {J.}~\bibnamefont
  {{Zhao}}}, \bibinfo {author} {\bibfnamefont {R.}~\bibnamefont {{Yu}}},
  \bibinfo {author} {\bibfnamefont {H.}~\bibnamefont {{Weng}}}, \ and\ \bibinfo
  {author} {\bibfnamefont {Z.}~\bibnamefont {{Fang}}},\ }\href@noop {}
  {\bibfield  {journal} {\bibinfo  {journal} {ArXiv e-prints}\ } (\bibinfo
  {year} {2015})},\ \Eprint {http://arxiv.org/abs/1511.05704} {arXiv:1511.05704
  [cond-mat.mtrl-sci]} \BibitemShut {NoStop}%
\bibitem [{\citenamefont {Volovik}(2013)}]{Volovik_2011}%
  \BibitemOpen
  \bibfield  {author} {\bibinfo {author} {\bibfnamefont {G.~E.}\ \bibnamefont
  {Volovik}},\ }\href@noop {} {\bibfield  {journal} {\bibinfo  {journal}
  {Analogue Gravity Phenomenology, Lecture Notes in Physics}\ }\textbf
  {\bibinfo {volume} {870}},\ \bibinfo {pages} {343} (\bibinfo {year}
  {2013})}\BibitemShut {NoStop}%
\bibitem [{\citenamefont {Rhim}\ and\ \citenamefont {Kim}(2015)}]{NLS_LL_PRB}%
  \BibitemOpen
  \bibfield  {author} {\bibinfo {author} {\bibfnamefont {J.-W.}\ \bibnamefont
  {Rhim}}\ and\ \bibinfo {author} {\bibfnamefont {Y.~B.}\ \bibnamefont {Kim}},\
  }\href {\doibase 10.1103/PhysRevB.92.045126} {\bibfield  {journal} {\bibinfo
  {journal} {Phys. Rev. B}\ }\textbf {\bibinfo {volume} {92}},\ \bibinfo
  {pages} {045126} (\bibinfo {year} {2015})}\BibitemShut {NoStop}%
\bibitem [{\citenamefont {Huh}\ \emph {et~al.}(2016)\citenamefont {Huh},
  \citenamefont {Moon},\ and\ \citenamefont
  {Kim}}]{NLS_long_range_Coulomb_interaction}%
  \BibitemOpen
  \bibfield  {author} {\bibinfo {author} {\bibfnamefont {Y.}~\bibnamefont
  {Huh}}, \bibinfo {author} {\bibfnamefont {E.-G.}\ \bibnamefont {Moon}}, \
  and\ \bibinfo {author} {\bibfnamefont {Y.~B.}\ \bibnamefont {Kim}},\ }\href
  {\doibase 10.1103/PhysRevB.93.035138} {\bibfield  {journal} {\bibinfo
  {journal} {Phys. Rev. B}\ }\textbf {\bibinfo {volume} {93}},\ \bibinfo
  {pages} {035138} (\bibinfo {year} {2016})}\BibitemShut {NoStop}%
\bibitem [{\citenamefont {Yan}\ \emph {et~al.}(2016)\citenamefont {Yan},
  \citenamefont {Huang},\ and\ \citenamefont
  {Wang}}]{NLS_Collective_modes_2015}%
  \BibitemOpen
  \bibfield  {author} {\bibinfo {author} {\bibfnamefont {Z.}~\bibnamefont
  {Yan}}, \bibinfo {author} {\bibfnamefont {P.-W.}\ \bibnamefont {Huang}}, \
  and\ \bibinfo {author} {\bibfnamefont {Z.}~\bibnamefont {Wang}},\ }\href
  {\doibase 10.1103/PhysRevB.93.085138} {\bibfield  {journal} {\bibinfo
  {journal} {Phys. Rev. B}\ }\textbf {\bibinfo {volume} {93}},\ \bibinfo
  {pages} {085138} (\bibinfo {year} {2016})}\BibitemShut {NoStop}%
\bibitem [{\citenamefont {Kopnin}\ \emph {et~al.}(2011)\citenamefont {Kopnin},
  \citenamefont {Heikkila},\ and\ \citenamefont
  {Volovik}}]{Flatband_HTC_Heikkila_2011PRB}%
  \BibitemOpen
  \bibfield  {author} {\bibinfo {author} {\bibfnamefont {N.~B.}\ \bibnamefont
  {Kopnin}}, \bibinfo {author} {\bibfnamefont {T.~T.}\ \bibnamefont
  {Heikkila}}, \ and\ \bibinfo {author} {\bibfnamefont {G.~E.}\ \bibnamefont
  {Volovik}},\ }\href {\doibase 10.1103/PhysRevB.83.220503} {\bibfield
  {journal} {\bibinfo  {journal} {Phys. Rev. B}\ }\textbf {\bibinfo {volume}
  {83}},\ \bibinfo {pages} {220503} (\bibinfo {year} {2011})}\BibitemShut
  {NoStop}%
\bibitem [{\citenamefont {Volovik}(2015)}]{Flatband_HTC_Volovik_2014arXiv}%
  \BibitemOpen
  \bibfield  {author} {\bibinfo {author} {\bibfnamefont {G.~E.}\ \bibnamefont
  {Volovik}},\ }\href {http://stacks.iop.org/1402-4896/2015/i=T164/a=014014}
  {\bibfield  {journal} {\bibinfo  {journal} {Physica Scripta}\ }\textbf
  {\bibinfo {volume} {2015}},\ \bibinfo {pages} {014014} (\bibinfo {year}
  {2015})}\BibitemShut {NoStop}%
\bibitem [{\citenamefont {{Heikkila}}\ and\ \citenamefont
  {{Volovik}}(2015{\natexlab{b}})}]{Flatband_HTC_Heikkila_2015arXiv}%
  \BibitemOpen
  \bibfield  {author} {\bibinfo {author} {\bibfnamefont {T.~T.}\ \bibnamefont
  {{Heikkila}}}\ and\ \bibinfo {author} {\bibfnamefont {G.~E.}\ \bibnamefont
  {{Volovik}}},\ }\href@noop {} {\bibfield  {journal} {\bibinfo  {journal}
  {ArXiv e-prints}\ } (\bibinfo {year} {2015}{\natexlab{b}})},\ \Eprint
  {http://arxiv.org/abs/1504.05824} {arXiv:1504.05824 [cond-mat.mtrl-sci]}
  \BibitemShut {NoStop}%
\bibitem [{\citenamefont {Dahlmann}\ and\ \citenamefont
  {Schnering}(1973)}]{CaP3}%
  \BibitemOpen
  \bibfield  {author} {\bibinfo {author} {\bibfnamefont {W.}~\bibnamefont
  {Dahlmann}}\ and\ \bibinfo {author} {\bibfnamefont {H.~V.}\ \bibnamefont
  {Schnering}},\ }\href@noop {} {\bibfield  {journal} {\bibinfo  {journal}
  {Naturwissenschaften}\ }\textbf {\bibinfo {volume} {60}},\ \bibinfo {pages}
  {518} (\bibinfo {year} {1973})}\BibitemShut {NoStop}%
\bibitem [{\citenamefont {Bauhofer}\ \emph {et~al.}(1981)\citenamefont
  {Bauhofer}, \citenamefont {Wittmann},\ and\ \citenamefont
  {Schnering}}]{CaAs3_BaAs3}%
  \BibitemOpen
  \bibfield  {author} {\bibinfo {author} {\bibfnamefont {W.}~\bibnamefont
  {Bauhofer}}, \bibinfo {author} {\bibfnamefont {M.}~\bibnamefont {Wittmann}},
  \ and\ \bibinfo {author} {\bibfnamefont {H.}~\bibnamefont {Schnering}},\
  }\href {\doibase http://dx.doi.org/10.1016/0022-3697(81)90122-0} {\bibfield
  {journal} {\bibinfo  {journal} {Journal of Physics and Chemistry of Solids}\
  }\textbf {\bibinfo {volume} {42}},\ \bibinfo {pages} {687 } (\bibinfo {year}
  {1981})}\BibitemShut {NoStop}%
\bibitem [{\citenamefont {Chen}\ \emph {et~al.}(2003)\citenamefont {Chen},
  \citenamefont {Zhu},\ and\ \citenamefont {Yamanaka}}]{SrP3_Chen2003449}%
  \BibitemOpen
  \bibfield  {author} {\bibinfo {author} {\bibfnamefont {X.}~\bibnamefont
  {Chen}}, \bibinfo {author} {\bibfnamefont {L.}~\bibnamefont {Zhu}}, \ and\
  \bibinfo {author} {\bibfnamefont {S.}~\bibnamefont {Yamanaka}},\ }\href
  {\doibase http://dx.doi.org/10.1016/S0022-4596(03)00142-7} {\bibfield
  {journal} {\bibinfo  {journal} {Journal of Solid State Chemistry}\ }\textbf
  {\bibinfo {volume} {173}},\ \bibinfo {pages} {449 } (\bibinfo {year}
  {2003})}\BibitemShut {NoStop}%
\bibitem [{\citenamefont {Deller}\ and\ \citenamefont
  {Eisenmann}(1976)}]{SrAs3}%
  \BibitemOpen
  \bibfield  {author} {\bibinfo {author} {\bibfnamefont {K.}~\bibnamefont
  {Deller}}\ and\ \bibinfo {author} {\bibfnamefont {B.}~\bibnamefont
  {Eisenmann}},\ }\href@noop {} {\bibfield  {journal} {\bibinfo  {journal}
  {Journal for Nature Research, Part B : Inorganic Chemistry, Organic
  Chemistry}\ }\textbf {\bibinfo {volume} {31}},\ \bibinfo {pages} {1550}
  (\bibinfo {year} {1976})}\BibitemShut {NoStop}%
\bibitem [{\citenamefont {Perdew}\ \emph {et~al.}(1996)\citenamefont {Perdew},
  \citenamefont {Burke},\ and\ \citenamefont {Ernzerhof}}]{Perdew:1996iq}%
  \BibitemOpen
  \bibfield  {author} {\bibinfo {author} {\bibfnamefont {J.~P.}\ \bibnamefont
  {Perdew}}, \bibinfo {author} {\bibfnamefont {K.}~\bibnamefont {Burke}}, \
  and\ \bibinfo {author} {\bibfnamefont {M.}~\bibnamefont {Ernzerhof}},\ }\href
  {\doibase 10.1103/PhysRevLett.77.3865} {\bibfield  {journal} {\bibinfo
  {journal} {Phys. Rev. Lett.}\ }\textbf {\bibinfo {volume} {77}},\ \bibinfo
  {pages} {3865} (\bibinfo {year} {1996})}\BibitemShut {NoStop}%
\bibitem [{\citenamefont {Bl\"ochl}(1994)}]{Blochl:1994zz}%
  \BibitemOpen
  \bibfield  {author} {\bibinfo {author} {\bibfnamefont {P.~E.}\ \bibnamefont
  {Bl\"ochl}},\ }\href {\doibase 10.1103/PhysRevB.50.17953} {\bibfield
  {journal} {\bibinfo  {journal} {Phys. Rev. B}\ }\textbf {\bibinfo {volume}
  {50}},\ \bibinfo {pages} {17953} (\bibinfo {year} {1994})}\BibitemShut
  {NoStop}%
\bibitem [{\citenamefont {Mostofi}\ \emph {et~al.}(2008)\citenamefont
  {Mostofi}, \citenamefont {Yates}, \citenamefont {Lee}, \citenamefont {Souza},
  \citenamefont {Vanderbilt},\ and\ \citenamefont {Marzari}}]{Mostofi:2008ff}%
  \BibitemOpen
  \bibfield  {author} {\bibinfo {author} {\bibfnamefont {A.~A.}\ \bibnamefont
  {Mostofi}}, \bibinfo {author} {\bibfnamefont {J.~R.}\ \bibnamefont {Yates}},
  \bibinfo {author} {\bibfnamefont {Y.-S.}\ \bibnamefont {Lee}}, \bibinfo
  {author} {\bibfnamefont {I.}~\bibnamefont {Souza}}, \bibinfo {author}
  {\bibfnamefont {D.}~\bibnamefont {Vanderbilt}}, \ and\ \bibinfo {author}
  {\bibfnamefont {N.}~\bibnamefont {Marzari}},\ }\href {\doibase
  http://dx.doi.org/10.1016/j.cpc.2007.11.016} {\bibfield  {journal} {\bibinfo
  {journal} {Comput. Phys. Commun.}\ }\textbf {\bibinfo {volume} {178}},\
  \bibinfo {pages} {685 } (\bibinfo {year} {2008})}\BibitemShut {NoStop}%
\bibitem [{\citenamefont {Marzari}\ \emph {et~al.}(2012)\citenamefont
  {Marzari}, \citenamefont {Mostofi}, \citenamefont {Yates}, \citenamefont
  {Souza},\ and\ \citenamefont {Vanderbilt}}]{Marzari:2012eu}%
  \BibitemOpen
  \bibfield  {author} {\bibinfo {author} {\bibfnamefont {N.}~\bibnamefont
  {Marzari}}, \bibinfo {author} {\bibfnamefont {A.~A.}\ \bibnamefont
  {Mostofi}}, \bibinfo {author} {\bibfnamefont {J.~R.}\ \bibnamefont {Yates}},
  \bibinfo {author} {\bibfnamefont {I.}~\bibnamefont {Souza}}, \ and\ \bibinfo
  {author} {\bibfnamefont {D.}~\bibnamefont {Vanderbilt}},\ }\href {\doibase
  10.1103/RevModPhys.84.1419} {\bibfield  {journal} {\bibinfo  {journal} {Rev.
  Mod. Phys.}\ }\textbf {\bibinfo {volume} {84}},\ \bibinfo {pages} {1419}
  (\bibinfo {year} {2012})}\BibitemShut {NoStop}%
\bibitem [{\citenamefont {Mong}\ and\ \citenamefont
  {Shivamoggi}(2011)}]{surface_states_PhysRevB.83.125109}%
  \BibitemOpen
  \bibfield  {author} {\bibinfo {author} {\bibfnamefont {R.~S.~K.}\
  \bibnamefont {Mong}}\ and\ \bibinfo {author} {\bibfnamefont {V.}~\bibnamefont
  {Shivamoggi}},\ }\href {\doibase 10.1103/PhysRevB.83.125109} {\bibfield
  {journal} {\bibinfo  {journal} {Phys. Rev. B}\ }\textbf {\bibinfo {volume}
  {83}},\ \bibinfo {pages} {125109} (\bibinfo {year} {2011})}\BibitemShut
  {NoStop}%
\end{thebibliography}%

\end{document}